\documentclass[11pt,showpacs,floatfix,superscriptaddress,notitlepage]{revtex4-1}

\newcommand{\ket}[1]{| #1 \rangle}
\newcommand{\bra}[1]{\langle #1 |}

\newcommand{\vect}[1]{\mathbf{#1}}

\newcommand{\colvect}[2]{\begin{bmatrix} #1 \\ #2 \end{bmatrix}}

\usepackage{graphicx}
\usepackage{amssymb,amsmath}

\begin{document}

\author{N. Yokomizo}
\email{yokomizo@if.usp.br}
\affiliation{Universidade de S\~{a}o Paulo, Instituto de F\'{i}sica\\
CP 66318, CEP 05315-970 S\~{a}o Paulo, SP, Brazil}

\title{Radiation from electrons in graphene in strong electric field}
\date{\today}

\begin{abstract}
We study the interaction of electrons in graphene with the quantized electromagnetic field in the presence of an applied uniform electric field using the Dirac model of graphene. Electronic states are represented by exact solutions of the Dirac equation in the electric background, and amplitudes of first-order Feynman diagrams describing the interaction with the photon field are calculated for massive Dirac particles in both valleys. Photon emission probabilities from a single electron and from a many-electron system at the charge neutrality point are derived, including the angular and frequency dependence, and several limiting cases are analyzed. The pattern of photon emission at the Dirac point in a strong field is determined by an interplay between the nonperturbative creation of electron-hole pairs and spontaneous emission, allowing for the possibility of observing the Schwinger effect in measurements of the radiation emitted by pristine graphene under DC voltage.
\end{abstract}
\maketitle

\section{Introduction}

In the Dirac model of graphene, electrons propagating in the two-dimensional structure of the material are described by massless Dirac fermions, with the speed of light replaced by the Fermi velocity $v_F \simeq 10^{6} \text{ m/s}$ \cite{semenoff,castroneto,peres,katsnelson}. The electronic properties of the material correspond to those of a two-dimensional gas of massless relativistic particles \cite{novoselov}, and its unique transport and optoelectronic properties can be understood as manifestations of specific quantum electrodynamic effects \cite{katsnelson-07,shytov,gusynin,vassilevich}. The fine structure constant in the Dirac model of graphene is much larger than in QED, however, and relativistic effects are greatly enhanced in the condensed matter context, allowing for exotic predictions of QED to be tested in table-top experiments. Outstanding examples are the observation of Klein tunneling in graphene heterojunctions \cite{goldhaber,kim-09} and of an anomalous integer quantum Hall effect due to the linear dispersion relation of the massless Dirac particles \cite{novoselov,kim-05}.

A long standing prediction of quantum electrodynamics is the instability of its vacuum in the presence of a strong electric field \cite{greiner,FGS,ruffini}. If the field strength is larger than a certain critical value, electron-positron pairs can be created from the vacuum by the applied field in a vacuum decay process known as the Schwinger effect \cite{schwinger}. This nonperturbative effect cannot be observed in high energy experiments as the required critical electric field is inaccessible in the laboratory. The analogous effect in graphene is the nonperturbative creation of electron-hole pairs by an external electric field in a sample with Fermi energy at the Dirac point. For a supercritical Coulomb field, pair creation is signaled by the existence of ``atomic collapse'' states \cite{pereira,shytov-2,shytov-3}, which have been recently observed around impurities in graphene \cite{wang}. In a uniform electric field, the created pairs are accelerated by the field and contribute to the dc conductivity \cite{lewkowicz-09,rosenstein-81,dora,rosenstein-82,vandecasteele,GGY}. 

The nonperturbative creation of electron-hole pairs in a strong uniform electric field $E$ gives rise to a superlinear current-voltage $I$--$V$ characteristic with $I \propto E^{3/2}$ \cite{rosenstein-81,dora}, which corresponds to a direct manifestation of the Schwinger pair creation rate in two dimensions \cite{allor,GG95}. Superlinear $I$--$V$ characteristics were observed in low-mobility graphene samples near the Dirac point, with a transition to a linear regime in high-mobility samples, and this behavior was consistently interpreted in terms of an interplay between pair creation and defect scattering \cite{vandecasteele}, but a more realistic model is required for a detailed quantitative analysis. A clear observation of the Schwinger mechanism in a uniform electric field is thus still missing.

In this work, we study electron-photon interactions in graphene in the presence of a strong uniform electric field using the Dirac model. Our purpose is to provide a complete description of radiative processes in graphene in the regime where pair creation is relevant. The angular and frequency distribution of photon emission is a much more sensitive probe of the distribution of electrons and holes than the dc current, and may provide an alternative means of observing the Schwinger effect in graphene, as suggested in \cite{lewkowicz-84}. Moreover, electron-hole annihilation is a source of dissipation present even in pristine graphene, and may not be negligible in clean samples at low temperature, as indicated by the results of \cite{mecklenburg}. A detailed description of this process is thus required for the investigation of the dc conductivity in large clean graphene samples.

Photon emission by free electron-hole recombination in graphene was studied for a constant number of electrons and holes in \cite{mecklenburg}, and for a time-dependent number of pairs due to pair creation in a constant electric field in \cite{lewkowicz-84}. In these works, the photon emission rate was calculated by a straightforward application of Fermi's golden rule. In a strong electric field the energy of the electrons is not conserved, however, and Fermi's rule cannot be applied. An alternative method is thus required. Here we apply standard techniques of QED with unstable vacuum \cite{FGS} to address this problem. The study of radiative processes in an electromagnetic background has an extensive bibliography (see \cite{FGS} for references), and we adapt some of these results to the context of graphene physics. The studies of quantum processes in a constant electric field developed by Nikishov around the 1970s are of particular relevance for our purposes \cite{nikishov-69,nikishov-70,nikishov-71}. In our approach, the electric background is treated nonperturbatively, and the interaction with the quantized electromagnetic field is considered to first order in perturbation theory. Exact solutions of the Dirac equation in a uniform electric field are used as unperturbed states, and the amplitudes of Feynman graphs describing single photon emission are computed at tree level. These amplitudes, together with Bogoliubov coefficients which describe pair creation, are the elementary blocks in the description of arbitrary first-order radiative process from many-electron states in graphene.

The paper is organized as follows. We describe the general settings of the problem and review properties of exact solutions of the Dirac equation in a uniform electric field in Section \ref{sec:setting}. Amplitudes of first-order Feynman graphs associated with radiative processes are computed in Section \ref{sec:electron-photon}. The radiation emitted by a single electron in graphene is studied in Section \ref{sec:single}. This example allows us to illustrate the most relevant features of our approach in a simple context. The many-body problem is considered in full detail in Section \ref{sec:many-body}. A general framework for the calculation of first-order radiative processes in a constant electric field near the Dirac point is presented, and applied to the derivation of the photon emission rate at the Dirac point. Footprints of the Schwinger effect in the photon emission rate are discussed at the end of this section. An Appendix is included containing a review of the techniques employed in the integration of the required first-order amplitudes.

\section{Electrons in graphene}
\label{sec:setting}

In the Dirac model of graphene, low-energy electronic excitations are described by Dirac fields $\psi(x)$ in two spatial dimensions. The interaction with the quantized electromagnetic field, however, involves the emission and absorption of photons which travel in the physical three-dimensional space. In this section, we model this interaction, taking into account exactly the presence of an external classical electric field, and the mixed dimensionality of the system.

\subsection{The Dirac model}

Low-energy electronic excitations $\psi(t,\vec{x})$ in graphene at zero temperature and chemical potential (i.e. at the charge neutrality point) are well described by a Dirac equation in a $(2+1)$-dimensional Minkowski space,
\begin{equation}
(\gamma^\mu p_\mu - m v_F ) \psi(t,\vec{x}) = 0 \, , \quad  p_\mu = i \hbar \partial_\mu \, ,
\label{eq:dirac-equation}
\end{equation}
where $\psi(t,\vec{x})$ is a two-component spinorial field, and the $\gamma$-matrices satisfy the usual anti-commutation relations $\{\gamma^\mu, \gamma^\nu\} = 2 \eta^{\mu \nu}$. The metric is $\eta_{\mu \nu}=\text{diag}\, (+1,-1,-1)$ in coordinates $x^\mu=(v_F t,x^1,x^2)$, where $v_F \simeq 10^6 \text{m/s}$ is the speed of electrons in graphene, the analogue of the speed of light $c$ in this model. We let Greek indices run from $0$ to $2$, and represent vectors in 2D space with arrows. Boldface symbols will be reserved for vectors in 3D space. In the presence of an external electromagnetic potential field $A_\mu(x)$, the Dirac equation \eqref{eq:dirac-equation} is modified by the introduction of a minimal coupling substitution \cite{gusynin},
\begin{equation}
p_\mu \rightarrow P_\mu = p_\mu - \frac{e}{c} A_\mu \, ,
\label{eq:minimal-coupling}
\end{equation}
where $e$ is the charge of the electron.

There are two fermion species $\psi(t,\vec{x})$ in the Dirac model of graphene, corresponding to excitations about the distinct Dirac points in the Brillouin zone of graphene. The algebra of 
$\gamma $-matrices has two inequivalent representations in $(2+1)$-dimensions, and a distinct (pseudo spin) representation is associated with each Dirac point. These can be written explicitly in the form
\begin{equation}
\gamma^0 = \sigma_3 \, , \; \gamma^1 = i \sigma_2 \, , \; \gamma^2 = -i \kappa \sigma_1 \, ,  
\label{eq:gamma-representations}
\end{equation}
where the $\sigma_i$ are Pauli matrices, and $\kappa= \pm 1$ labels inequivalent representations. Inserting the representation \eqref{eq:gamma-representations} of the $\gamma$-matrices in \eqref{eq:dirac-equation}, the Dirac equation can be cast in Hamiltonian form,
\begin{gather}
i \hbar \, \partial_t \psi(t,\vec{x}) = H \psi(t,\vec{x}) \, , 
\nonumber \\
H = v_F \left( \sigma_1 p^1 + \kappa \sigma_2 p^2 + \sigma_3 m v_F \right) \, .
\label{eq:dirac-eq-hamiltonian}
\end{gather}

The Dirac fields $\psi(t,\vec x)$ are associated with three-dimensional Schr\"odinger wavefunctions
\begin{equation}
\phi_a(t,\vect x) = \psi_a(t,\vec{x}) \, \textrm{e}^{i \vec K_\kappa \cdot \vec x} \, f(z) \, \textrm{e}^{i p^3 z} \, , 
\label{eq:2d-3d-map}
\end{equation}
where $\vec K_\kappa$ is the quasimomentum of the corresponding Dirac point in reciprocal space, $a=1,2$ labels projections on the sublattices of the honeycomb lattice, and the function $f(z)$ represents the width of the material. A detailed description of $f(z)$ is not necessary for our purposes, except for the fact that it decays rapidly outside the $xy$-plane, and is normalized according to $\int dz |f(z)|^2 = 1$. We allow the graphene sheet to have a global momentum $p^3$ along the z axis, in order to account for the possibility of momentum transfer in this direction to some external system.

We are interested in the situation where the electromagnetic potential can be decomposed into two contributions,
\begin{equation}
A(t,\vect x) = \hat{A}(t,\vect x) + A^{ext}(t, \vect x) \, ,
\label{eq:A-decomp}
\end{equation}
where $\hat{A}(t,\vect x)$ is the quantized electromagnetic potential and $A^{ext}(t,\vect x)$ is the potential of a classical electromagnetic background (which will be described in the next section). This representation has the appropriate form for the study of photon emission (and absorption) in the presence of a classical electromagnetic field. Using \eqref{eq:A-decomp}, the minimal coupling prescription introduces two new terms in the Hamiltonian \eqref{eq:dirac-eq-hamiltonian}, which represent the interaction with photons and the action of the external classical field. The effect of the applied field will be taken into account exactly, meaning that we will take exact solutions of the Dirac equation in the external field as unperturbed states. The interaction term with the quantized field will be treated as a perturbation describing quantum processes of emission and absorption of photons in the presence of the electromagnetic background.

In the Coulomb gauge, the vector potential of the quantized electromagnetic field is
\begin{equation}
\vect A(t, \vect x) = c \sum_{\vect k \alpha} \sqrt{\frac{2 \pi \hbar}{V \omega}} \boldsymbol{\epsilon}_{\vect k \alpha} \left[ c_{\vect k \alpha} \, \textrm{e}^{i(\vect k \cdot \vect x - \omega t)} + c^\dagger_{\vect k \alpha} \, \textrm{e}^{- i(\vect k \cdot \vect x - \omega t)} \right] \, ,
\label{eq:vector-potential}
\end{equation}
where $\alpha=1,2$ is a polarization index, the $\boldsymbol{\epsilon}_{\vect k \alpha}$ are unit polarization vectors transversal to each other and to the momentum $\vect k$ satisfying the condition $\boldsymbol{\epsilon}_{\vect k \alpha} = \boldsymbol{\epsilon}_{(-\vect k) \alpha}$, and $c_{\vect k \alpha},c^\dagger_{\vect k \alpha}$ are annihilation and creation operators of photons. $V$ is the volume of the box regularization. From \eqref{eq:dirac-eq-hamiltonian}, we obtain a minimal coupling perturbation term
\begin{equation}
V_{int} = - e v_F \sqrt{\frac{2 \pi \hbar}{V \omega}} \sum_{\vect k \alpha} \vec \sigma \cdot \vec \epsilon_{\vect k \alpha} \left[ c_{\vect k \alpha} \, \textrm{e}^{i(\vect k \cdot \vect x - \omega t)} + c^\dagger_{\vect k \alpha} \, \textrm{e}^{- i(\vect k \cdot \vect x - \omega t)} \right]
\label{eq:perturbation}
\end{equation}
for $\kappa=1$. The same formula is valid for $\kappa=-1$ with the substitution $\vec \sigma \to \vec \sigma^*$. Notice that the scalar product $\vec \sigma \cdot \vec \epsilon_{\vect k \alpha}$ involves only 2D spatial vectors. Therefore, only the two first components of the polarization vectors couple to the spinor field. One-particle states of the quantized electromagnetic field will be represented in the basis $\ket{\vect k, \alpha}=c^\dagger_{\vect k \alpha} \ket{0}$. States with many photons are finite-norm superpositions of symmetrized products of such one-particle states. We denote the Fock space of photon states by $\mathcal{H}_{ph}$.

\subsection{Exact solutions in a constant electric field}
\label{sec:settings}

Let $A^{ext}(t, \vect x)$ be the potential of a constant uniform electric field parallel to the $xy$-plane of the graphene sample. We choose coordinates such that the electric field in the $xy$-plane is $\vect E(t,\vec x) = (E,0,0)$, with $eE >0$, and a gauge in which the electromagnetic potential is uniform and has only one nonzero component, $A_\mu(t, \vec x)= (0,Ect,0)$.

Writing the spinor field in the form
\begin{equation}
\psi(x) = (\gamma^{\mu} P_{\mu} + m v_F) \phi (x) \, ,  
\label{eq:psi-phi}
\end{equation}
where $\phi(x)$ is a new two-component spinor field, it follows from the Dirac equation that
\begin{equation}
\left[ P^2 - m^2 v_F^2 - i \frac{e E \hbar}{v_F} \gamma^0 \gamma^1 \right] \phi(x)=0 \, .
\label{eq:dirac-phi}
\end{equation}
This equation can be solved by separation of variables. Introducing orthonormalized spinors
\[
u_+ = \frac{1}{\sqrt{2}} \colvect{1}{1} \, , \qquad u_- = \frac{1}{\sqrt{2}} \colvect{1}{-1} \, , 
\] 
defined (up to a phase) by the relations
\[
\gamma^0 \gamma^1 u_s = s u_s \, , \quad s=\pm 1 \, ,
\]
and representing the spinors $\phi (x)$ in the factorized form
\begin{equation}
\phi _{\vec{p},s}(t,\vec{x})=e^{i\vec{p}\cdot \vec{x} / \hbar} \, \varphi_{\vec{p},s}(t) \, u_s \, ,
\label{eq:phi-varphi}
\end{equation}
where $\varphi_{\vec{p},s}$ is a scalar, we obtain the equation of motion:
\begin{equation}
\left\{ \frac{d^2}{dt^2} + \left(\frac{v_F}{\hbar}\right)^2 P^2(t)  + i \frac{v_F}{\hbar} e E s \right\} \varphi _{\vec{p},s}(t)=0 \, ,
\label{eq:time-evolution}
\end{equation}
where
\begin{equation}
P(t) = \sqrt{(p_1-eEt)^2 + p_2^2 + m^2 v_F^2} \, .
\label{eq:asymptotic-momentum}
\end{equation}
The Eq.~\eqref{eq:time-evolution} has the form of a Weber equation, with solutions given by parabolic cylinder (Weber) functions [see \cite{erdelyi}].

We are interested in two special complete sets of solutions, the so-called in- and out-solutions, which we represent as $_{\pm}\psi_{\vec p}(x), ^{\pm}\psi_{\vec p}(x)$, respectively. Such states are characterized by the asymptotic conditions:
\begin{align}
i \hbar \partial_0[ _{\pm}\psi_{\vec p}(x) ] &\simeq \pm v_F \, P(t) \,  _{\pm}\psi_{\vec p}(x) \, , &  \text{for } t &\to - \infty \, , 
\notag \\
i \hbar \partial_0[ ^{\pm}\psi_{\vec p}(x) ] &\simeq \pm v_F \, P(t) \,  ^{\pm}\psi_{\vec p}(x) \, , & \text{for } t &\to \infty \, .
\label{eq:asymptotic-conditions}
\end{align}
An in-state $_{\pm}\psi_{\vec p}(x)$ describes a particle with momentum $\vec P(t)=\big(p^1+eEt,p^2\big)$ and a well-defined energy sign $\pm$ at the distant past. Similarly, an out-state $^{\pm}\psi_{\vec p}(x)$ describes a particle with momentum $\vec P(t)$ and a well-defined energy sign $\pm$ at the distant future (see details in \cite{GG95}).

Expressions for the in- and out-states satisfying the asymptotic conditions \eqref{eq:asymptotic-conditions} in arbitrary dimensions were given in \cite{GG95}. In terms of the variables:
\begin{gather}
\lambda  = \frac{1}{eE} \left[ \frac{v_F}{\hbar} p_2^2 + \frac{m^2 v_F^3}{\hbar} \right] \, , \qquad  \nu  = \frac{i \lambda}{2} \, , \nonumber \\
\xi(t)  = \sqrt{\frac{v_F}{\hbar}} \frac{eE t - p_1}{\sqrt{eE}} \, ,
\label{eq:new-parameters}
\end{gather}
the asymptotic states read:
\begin{align}
_{\pm}\psi_{\vec p}(x) &= (\gamma P + m v_F) \textrm{e}^{i \vec p \cdot \vec x / \hbar} \, {} _{\pm}\varphi_{\vec p \pm}(t) u_\pm \, , \nonumber \\
^{\pm}\psi_{\vec p}(x) &= (\gamma P + m v_F) \textrm{e}^{i \vec p \cdot \vec x / \hbar} \, {} ^{\pm}\varphi_{\vec p \mp}(t) u_\mp \, ,
\label{eq:cte-field-solutions}
\end{align}
where
\begin{gather}
_{+} ^{-}\varphi_{\vec p s}(t) = C D_{\nu - \frac{1+s}{2}}[\pm (1-i) \xi] \, , \nonumber \\
_{-} ^{+}\varphi_{\vec p s}(t) = C D_{-\nu - \frac{1-s}{2}}[\pm (1+i) \xi] \, ,
\label{eq:varphi-modes}
\end{gather}
are solutions of Eq.~\eqref{eq:time-evolution}, and the normalization constant is
\begin{equation}
C = \frac{1}{\sqrt{S}} \frac{\textrm{e}^{-\pi \lambda/8}}{\sqrt{2}A} \, ,
\label{eq:C-def}
\end{equation}
with $A=\sqrt{eE\hbar/v_F}$. The $D$'s are parabolic cylinder functions \cite{erdelyi}, and $S$ is the area of the two-dimensional space in a box normalization.

We can express these solutions in a more explicit manner. Inserting \eqref{eq:varphi-modes} in \eqref{eq:cte-field-solutions}, and computing the derivatives, we find:
\begin{align}
^-_+\psi_{\vec p}(x) &= C \big\{ (mv_F - i \kappa p_2) D_{\nu-1}[\pm(1-i)\xi] u_+  \mp  A (1+i) D_\nu[\pm(1-i)\xi] u_- \big\} \textrm{e}^{i \vec p \cdot \vec x / \hbar} \, ,
\label{eq:asymptotic-states-dplus} \\
^+_-\psi_{\vec p}(x) &= C \big\{ (mv_F + i \kappa p_2) D_{-\nu-1}[\pm(1+i)\xi] u_- \pm  A (1-i) D_{-\nu}[\pm(1+i)\xi] u_+ \big\} \textrm{e}^{i \vec p \cdot \vec x / \hbar} \, .
\label{eq:asymptotic-states-dminus}
\end{align}
This is the representation which will be employed in the calculation of the amplitudes of the Feynman graphs. The three-dimensional wavefunctions associated with such asymptotic states are obtained by applying the prescription \eqref{eq:2d-3d-map} to the spinor fields in Eqs.~\eqref{eq:asymptotic-states-dplus} and \eqref{eq:asymptotic-states-dminus}, and will be represented as
\begin{equation}
_{\pm}\chi_{\vec p}(t, \vect x) = {} _{\pm}\psi_{\vec p}(t, \vec x) \, \textrm{e}^{i \vec K_\kappa \cdot \vec x} \, f(z) \, \textrm{e}^{i k_z z} \, , 
\label{eq:asymptotic-3d}
\end{equation}
and similarly for the out-states.

Since the sets $\{_{\pm}\psi_{\vec p}(x)\}$ and $\{^{\pm}\psi_{\vec p}(x)\}$ of in- and out-solutions are both complete in the space of solutions of the Dirac equation, it is possible to expand an out-solution as a superposition of in-solutions, and vice versa. Therefore, we can define coefficients $g$ through the relations:
\begin{align}
^\zeta \psi_{\vec p} = g(_+ \mid ^\zeta) \, _{+} \psi_{\vec p} + g(_- \mid ^\zeta) \, _{-} \psi_{\vec p} \, , \nonumber \\
_\zeta \psi_{\vec p} = g(^+ \mid _\zeta) \, ^{+} \psi_{\vec p} + g(^- \mid _\zeta) \, ^{-} \psi_{\vec p} \, .
\label{eq:g-matrices}
\end{align}
The transformation above does not mix momenta, i.e., $\vec p$ is a constant of motion. Moreover, from the general theory discussed in \cite{FGS} it is always true that
\begin{equation}
g(_\eta \mid ^\zeta) = \overline{g(^\zeta \mid _\eta )}  \, .
\label{eq:g-cc}
\end{equation}
For a uniform electric field, the following identity also holds:
\begin{equation}
g(_\eta \mid ^\zeta)(- \vec p) = g(^{-\eta} \mid _{-\zeta})(\vec p) \, ,
\label{eq:g-t-rev}
\end{equation}
and it follows that all $g$-coefficients can be written in terms of $g(_- \mid ^+)$ and $g(_+ \mid ^+)$.

The in-state $_\zeta \psi_{\vec p}$ represents a particle with momentum $\vec P(t)$ and negative energy in the asymptotic past, but a superposition of positive and negative energy states with momentum $\vec P(t)$ in the asymptotic future, with amplitudes $g(^+ \mid _\zeta)$ and $g(^- \mid _\zeta)$, respectively, according to Eqs.~\eqref{eq:asymptotic-conditions} and \eqref{eq:g-matrices}. The transition between these asymptotic behaviors occurs during an interval of time of width $\Delta t_{bt} \sim \sqrt{\hbar/eEv_F} (1+\lambda)$ around $t_{bt} = p_1/eE$ \cite{nikishov-70,nikishov-71,GG95}. The longitudinal momentum of the particle is equal to zero at the band transition time, $P_1(t_{st})=0$.

The squared $g$-coefficients are given by
\begin{align}
|g(_- \mid ^+)|^2 &= |g(_+ \mid ^-)|^2 = \textrm{e}^{- \pi \lambda} \, , \notag \\
|g(_+ \mid ^+)|^2 &= |g(_- \mid ^-)|^2 = 1 - \textrm{e}^{- \pi \lambda} \, ,
\label{eq:pair-creation-prob}
\end{align}
and depend on $\vec p$ only through $p_2^2$. In the context of quantum field theory, $\exp(- \pi \lambda)$ describes the probability of creation of an electron-positron pair with quasimomentum $\pm \vec p$ by the electric field. In the Dirac model of graphene, it corresponds to the probability that an electron initially in the negative energy band tunnels to the upper energy band as a result of the action of the electric field.

\section{Electron-photon interactions in a uniform electric field}
\label{sec:electron-photon}

\subsection{Photon emission by electron in the upper band}

\subsubsection{Intraband transition}
\label{sec:upper-intra}

\begin{figure}
\begin{center}
\includegraphics[scale=1]{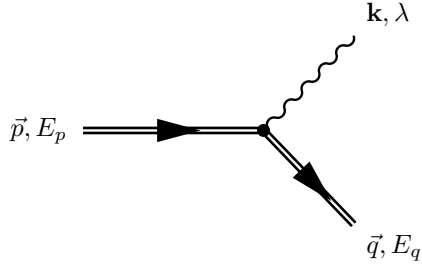}
\end{center}
\caption{Photon emission by an electron in graphene. An electron with quasimomentum $\vec p$ decays into a state with quasimomentum $\vec q$ due to the emission of a photon with wavenumber $\vect k$ and polarization $\alpha$. The electron is confined to the graphene plane, but the photon propagates in the physical three-dimensional space.}
\label{fig:feynman-graph}
\end{figure}

Consider the process where an electron with initial quasimomentum $\vec p$ and positive energy $E_p(t)>0$ (for $t\to -\infty$) emits a photon with wavenumber $\vect k$  and polarization $\boldsymbol{\epsilon}_{\vect k \alpha}$ in the presence of a uniform electric field with the potential $A_\mu$ described in Section \ref{sec:settings}, and after the interaction has quasimomentum $\vec q$ and energy $E_q(t)>0$ (for $t \to \infty$). This process corresponds to the emission of a single photon by a conduction electron (Fig.~\ref{fig:feynman-graph}). We will restrict to intra-valley scatterings throughout the paper.

To first order, the amplitude of the process is given by:
\begin{equation}
M^{(e)}_{\vect k \alpha} = \frac{1}{i\hbar} \int d\vect x \, dt \, \Bigl({} ^{+}\chi^\dagger_{\vec q} \otimes \bra{\vect k,\alpha} \Bigr) V_{int} \Bigl( {} _{+}\chi_{\vec p} \otimes \ket{0} \Bigr) \, .
\label{eq:M-e}
\end{equation}
From the definition of $V_{int}$ in Eq.~\eqref{eq:perturbation}, we obtain
\begin{equation}
M_{\vect k \alpha}^{(e)} = i e \frac{v_F}{\hbar} \sqrt{\frac{2 \pi \hbar}{V \omega}} \int dz \, |f(z)|^2 \textrm{e}^{iz(p_z - q_z - \hbar k_z)/\hbar} 
 \int d\vec x \, dt  {}\, ^{+}\psi^\dagger_{\vec q}  ( \vec \sigma \cdot \vec \epsilon_{\vect k \alpha} ) {} _{+}\psi_{\vec p} \, \textrm{e}^{-i (\vec k \cdot \vec x - \omega t)} \, .
\label{amplitude-M}
\end{equation}
The integral in  $z$ is approximately equal to one for low transverse momentum transfer, which we assume (in this case the exponential is approximated by the zeroth-order constant term, as in the usual dipole approximation). From the explicit expressions for the electronic states given in \eqref{eq:asymptotic-states-dplus} and \eqref{eq:asymptotic-states-dminus}, we can write:
\begin{equation}
M_{\vect k \alpha}^{(e)} = i e \frac{v_F}{\hbar} \sqrt{\frac{2 \pi \hbar}{V \omega}} \left[ \int d\vec x \, \textrm{e}^{i(\vec p -\hbar \vec k - \vec q)\cdot \vec x/\hbar} \right] C C^\prime N^{(e)}_{\vect k \alpha} \, ,
\label{amplitude-M-2}
\end{equation}
where $N^{(e)}_{\vect k \alpha}$ is the integral of the time-dependent factors in \eqref{amplitude-M}, and reads:
\begin{equation}
N^{(e)}_{\vect k \alpha} =  2i A^2 J^\ast_{00} \, S_\alpha^{+-} + (1+i) A \pi_2^\ast J^\ast_{10} \, S_\alpha^{++}
 + (1+i) A \chi_2^\ast J^\ast_{01} S_\alpha^ {--} + \pi_2^\ast \chi_2^\ast J^\ast_{11} S_{\alpha}^{-+} \, ,
\label{eq:amplitude-I}
\end{equation}
where $\alpha=1,2$ is a polarization index, $C$ and $C^\prime$ are defined in \eqref{eq:C-def} (primed quantities refer to the initial state, and unprimed quantities to the final state), and we have introduced new quasimomenta variables:
\begin{equation}
\pi_2 = mv_F + i \kappa p_2 \, , \qquad \chi_2 = mv_F + i \kappa q_2 \, .
\label{eq:pi-chi-def}
\end{equation}
The parameters $S$ encode the dependence on the polarization of the emitted photon,
\begin{equation}
S_\alpha^{rs} = u_r^\dagger (\sigma_1 \epsilon^1_{\vect k \alpha} + \kappa \sigma_2 \epsilon^2_{\vect k \alpha}) u_s \, , \qquad r,s=\pm \, ,
\label{eq:S-definition}
\end{equation}
while the $J$ functions correspond to the time integrals
\begin{equation}
J_{j'j}(\omega) = \int_{-\infty}^{+ \infty} dt \, D_{-i \lambda'/2-j'}[-(1+i)\xi'] D_{-i \lambda/2-j}[(1+i)\xi] \textrm{e}^{-i \omega t}\, ,
\label{eq:nikishov-J}
\end{equation}
with $\xi,\lambda$ given in \eqref{eq:new-parameters}. Except for a delta of quasimomentum conservation and simple factors, the amplitude $M_{\vect k \alpha}^{(e)}$ is determined by the coefficient $N^{(e)}_{\vect k \alpha}$ introduced in \eqref{eq:amplitude-I}.

The functions $S_\alpha^{rs}$ are easily calculated for a given choice of polarization vectors $\boldsymbol{\epsilon}_{\vect k \alpha}$. We adopt the convention used in \cite{mecklenburg},
\begin{align*}
\vect k & = (\sin \theta_\gamma \cos \phi_\gamma \, , \, \sin \theta_\gamma \sin \phi_\gamma \, , \, \cos \theta_\gamma) \, , 
\\
\boldsymbol \epsilon_{\vect k 1} & = (- \sin \phi_\gamma \, , \, \cos \phi_\gamma  \, , \, 0) \, ,
\\
\boldsymbol \epsilon_{\vect k 2} & = (-\cos \theta_\gamma \cos \phi_\gamma \, , \, -\cos \theta_\gamma \sin \phi_\gamma  \, , \, \sin \theta_\gamma ) \, ,
\end{align*}
for $\vect k$ in the upper spatial region, $k_z \geq 0$, and $\boldsymbol{\epsilon}_{\vect k \alpha} = \boldsymbol{\epsilon}_{(-\vect k) \alpha}$. The polarization $\boldsymbol{\epsilon}_{\vect k 1}$ is parallel to the graphene sample, while the polarization $\boldsymbol{\epsilon}_{\vect k 2}$ is orthogonal to $\boldsymbol{\epsilon}_{\vect k 1}$ and $\vect k$. We obtain:
\begin{align}
S_1^{--} = - S_1^{++} & = \sin \phi_\gamma \, , \notag \\
S_1^{+-} = - S_1^{-+} & = i \kappa \cos \phi_\gamma \, , \notag \\
S_2^{--} = - S_2^{++} & = \cos \theta_\gamma \cos \phi_\gamma \, , \notag \\
S_2^{+-} = - S_2^{-+} & = - i \kappa \cos \theta_\gamma \sin \phi_\gamma \, .
\label{eq:spin-S}
\end{align}

The $J_{j'j}$ integrals were computed in \cite{nikishov-71}, in a study of radiative processes in quantum electrodynamics in the presence of a constant electric field. The techniques for solving these integrals and similar ones which will be encountered in later sections are reviewed in detail in the Appendix. It tuns out that the dependence on the longitudinal momenta can be essentially factored out,
\begin{align}
& J_{j'j} = \frac{\hbar}{A v_F} \exp\left( - i \omega \frac{p_1 + q_1}{2eE} \right) \textrm{e}^{-i \beta \varphi} \textrm{e}^{(j'-j)\varphi} I_{j'j}(\rho) \, ,  \nonumber \\
& I_{j'j}(\rho) = \int_{-\infty}^{+\infty} dv D_{-i\lambda'/2 - j'}[(1+i)v] D_{-i\lambda/2-j}[-(1+i)v] \textrm{e}^{- i \rho v} \, ,
\label{eq:J-I}
\end{align}
where $\beta = (\lambda - \lambda')/2$, and $\rho,\varphi$ are hyperbolic coordinates defined through the transformation:
\begin{align}
A \, \rho \cosh \varphi & = - \omega \frac{\hbar}{v_F} \, , \nonumber \\
A \, \rho \sinh \varphi & = p_1-q_1 \, .
\label{eq:rho-phi}
\end{align}
Explicit formulas for the $I_{j'j}$ can be written in terms of confluent hypergeometric functions \cite{nikishov-71},
\begin{multline}
I_{j'j}(\rho) = \sqrt{\pi} \textrm{e}^{-i \frac{\pi}{4} } \textrm{e}^{i \rho^2/4}  \\ \times
\left\{   \frac{\Gamma(-i \beta -j+j')}{\Gamma(i \frac{\lambda'}{2} + j')}  \left( \frac{-i \rho}{\sqrt{2}}  \textrm{e}^{-i\frac{\pi}{4}} \right)^{i \beta + j-j'} \Phi\left( i \frac{\lambda}{2} + j, 1+i\beta +j - j'; -i\frac{\rho^2}{2} \right) \right. \\
+  \left. \frac{\Gamma(i \beta +j -j')}{\Gamma(i \frac{\lambda}{2} + j)} \left( \frac{i \rho}{\sqrt{2}}\textrm{e}^{-i\frac{\pi}{4}} \right)^{-i \beta - j+j'} \Phi\left( i \frac{\lambda'}{2} + j', 1 - i \beta-j+j'; -i\frac{\rho^2}{2} \right)  \right\} \, .
\label{eq:I-integrals}
\end{multline}
The $I_{j'j}$ integrals are not independent, because of contiguity relations holding among the confluent hypergeometric functions appearing in \eqref{eq:I-integrals}, which lead to (see proof in the Appendix):
\begin{align}
\textrm{e}^{i\pi/4} \frac{\rho}{\sqrt{2}} I_{01} & = I_{00} - i \frac{\lambda'}{2} I_{11} \, , \nonumber \\
\textrm{e}^{i \pi/4} \frac{\rho}{\sqrt{2}} I_{10} & = - I_{00} + i \frac{\lambda}{2} I_{11} \, .
\label{eq:contiguity-relations}
\end{align}
Therefore, only two of such integrals are independent, allowing us to write the amplitudes $N^{(e)}_{\vect k \alpha}$ in terms of $I_{00}$ and $I_{11}$ only.

Substituting Eqs.~\eqref{eq:spin-S}--\eqref{eq:contiguity-relations} in the formula for $N^{(e)}_{\vect k \alpha}$ given in \eqref{eq:amplitude-I}, we obtain:
\begin{multline}
N^{(e)}_{\vect k \alpha} = \frac{\hbar}{A v_F} \textrm{e}^{i \beta \varphi} \exp\left(i \omega \frac{p_1+q_1}{2eE} \right) \left\{I_{00}^\ast \left[ 2iA^2 S_\alpha^{+-} + 2i \frac{A}{\rho} \left( \pi_2^\ast \textrm{e}^{\varphi} + \chi_2^\ast \textrm{e}^{-\varphi} \right) S_\alpha^{--} \right]        \right. \\
\left. + I_{11}^\ast \left[ \pi_2^\ast \chi_2^\ast S_\alpha^{-+} - \frac{A}{\rho} \left( \lambda \pi_2^\ast \textrm{e}^{\varphi} + \lambda^\prime \chi_2^\ast \textrm{e}^{-\varphi} \right) S_\alpha^{--} \right] \right\} \, .
\label{eq:explicit-N-lambda}
\end{multline}
This completes the calculation of the amplitude of the process. The corresponding differential transition probability is given by:
\begin{equation}
d\Gamma_{\vect k \alpha}^{(e)} = \frac{e^2 v_F^2}{\hbar} \frac{2\pi}{V \omega}  (C C')^2 \left|N^{(e)}_{\vect k \alpha} \right|^2 \, S (2\pi)^2 \delta\left(\frac{\vec p}{\hbar} - \frac{\vec q}{\hbar} - \vec k\right) \frac{S d(\vec q/\hbar)}{(2\pi)^2} \, \frac{V d \vect k}{(2\pi)^3}\, .
\label{eq:transition-M}
\end{equation}
For unpolarized photon emission, $\big|N^{(e)}_{\vect k \alpha} \big|^2$ should be replaced in \eqref{eq:transition-M} with the sum of the squared amplitudes over the polarizations, which we write in the form:
\begin{equation}
N^2 = \sum_{\alpha=1}^2 \left|N^{(e)}_{{\vect k \alpha}}\right|^2 = f_0 |I_{00}|^2 + f_1 |I_{11}|^2 + 2 \mathrm{Re} [f_2 I_{00}^\ast I_{11}] \, ,
\label{eq:N-squared}
\end{equation}
where the functions $f_i$ are obtained from \eqref{eq:spin-S} and \eqref{eq:explicit-N-lambda},
\begin{align}
f_0 & = 4  \left(\frac{A \hbar}{v_F}\right)^2 \left\{ 1 + \left|\frac{\pi_2 \textrm{e}^\varphi + \chi_2 \textrm{e}^{-\varphi}}{A \rho} \right|^2 - \sin^2 \theta_\gamma \left| \sin \phi_\gamma - i  \kappa \cos \phi_\gamma \left(\frac{\pi_2 \textrm{e}^\varphi + \chi_2 \textrm{e}^{-\varphi}}{A \rho} \right) \right|^2 \right\} \, , 
\label{eq:f00-full}
\\
f_1 & = \lambda \lambda^\prime  \left(\frac{A \hbar}{v_F}\right)^2 \left\{ 1 + \left|\frac{\pi_2 \textrm{e}^{-\varphi} + \chi_2 \textrm{e}^\varphi}{A \rho} \right|^2 - \sin^2 \theta_\gamma \left| \sin \phi_\gamma + i \kappa \cos \phi_\gamma \left(\frac{\pi_2 \textrm{e}^{-\varphi} + \chi_2 \textrm{e}^\varphi}{A \rho} \right) \right|^2 \right\} \, ,  
\label{eq:f11-full}
\\
f_2 & = -2i \left(\frac{A \hbar}{v_F}\right)^2 \left\{ \frac{\pi_2 \chi_2}{A^2} (1-\sin^2 \theta_\gamma \sin^2 \phi_\gamma)+ i \kappa \frac{\pi_2 \lambda - \chi_2 \lambda'}{A\rho}\left(\textrm{e}^\varphi - \textrm{e}^{-\varphi} \right) \sin^2 \theta_\gamma \sin \phi_\gamma \cos \phi_\gamma  \right. \nonumber \\
& \quad \left.  + (1- \sin^2 \theta_\gamma \cos^2 \phi_\gamma) \left[ \frac{\lambda \lambda'}{\rho^2} (\textrm{e}^{2\varphi} + \textrm{e}^{-2\varphi}) + \frac{\pi_2 \chi_2^\ast \lambda + \pi_2^\ast \chi_2 \lambda'}{(A\rho)^2} \right] \right\} \, ,
\label{eq:f10-full}
\end{align}
Integrating \eqref{eq:transition-M} in $\vec q$-space and summing over polarizations, we obtain the unpolarized photon emission probability:
\begin{equation}
\frac{d\Gamma^{(e)}_\alpha}{d \Omega \, d\omega} = \frac{e^2}{\hbar c} \left(\frac{v_F}{c}\right)^2 \frac{\omega}{(2 \pi)^2}  (C C')^2 S^2 N^2 \, .
\label{eq:transition-probability-er}
\end{equation}

All formulas were written in a form valid for both Dirac points, which are labeled by the index $\kappa$, and for arbitrary values of the mass gap $m$. From \eqref{eq:f00-full}--\eqref{eq:f10-full}, $f_0$ and $f_1$ are independent of $\kappa$, but $f_2$ depends on $\kappa$. The coefficients $f_i$ for $\kappa=-1$ have the same numerical value as those for $\kappa=+1$ with $q_2 \to -q_2,p_2 \to -p_2,\phi_\gamma \to -\phi_\gamma$. Therefore, the amplitude of the process in one Dirac point corresponds to that of the mirrored process (through the $xz$-plane) in the other Dirac point. For massless particles, all $f_i$ are independent of $\kappa$, and it is not necessary to distinguish between Dirac points.

\subsubsection{Interband transition}
\label{sec:upper-inter}

Consider the case of an electron with initial quasimomentum $\vec p$ and energy $E_p(t)>0$ in the asymptotic past, and final quasimomentum $\vec q$ and energy $E_q(t)<0$ in the asymptotic future. This corresponds to an interband transition in which a single electron-hole pair is annihilated. The amplitude of the process is given to first order by:
\begin{equation}
M^{(pa)}_{\vect k \alpha} = \frac{1}{i\hbar} \int d\vect x \, dt \, \Bigl({} ^{-}\chi^\dagger_{\vec q} \otimes \bra{\vect k,\alpha} \Bigr) V_{int} \Bigl( {} _{+}\chi_{\vec p} \otimes \ket{0} \Bigr) \, .
\label{eq:pa-amplitude-1}
\end{equation}
Proceeding as in the last section, we find that:
\begin{equation}
M_{\vect k \alpha}^{(pa)} = i e \frac{v_F}{\hbar} \sqrt{\frac{2 \pi \hbar}{V \omega}} \left[ \int d\vec x \, \textrm{e}^{i(\vec p -\hbar \vec k - \vec q)\cdot \vec x/\hbar} \right] C C^\prime Q^{(pa)}_{\vect k \alpha} \, ,
\label{eq:pa-amplitude-2}
\end{equation}
where $Q^{(pa)}_{\vect k \alpha}$ is the integral of the time-dependent factors in Eq.~\eqref{eq:pa-amplitude-1},
\begin{equation}
Q^{(pa)}_{\vect k \alpha} = - 2 A^2 L^\ast_{00} \, S_\alpha^{--} - (1-i) A \pi_2^\ast  L^\ast_{10} \, S_\alpha^{-+} + (1+i) A \chi_2 L^\ast_{01} S_\alpha^ {+-} + \pi_2^\ast \chi_2 L^\ast_{11} S_{\alpha}^{++} \, .
\label{eq:amplitude-Q}
\end{equation}
The $L_{j'j}$'s are defined by:
\begin{equation}
L_{j'j} = \int_{-\infty}^{+ \infty} dt \, D_{-i \lambda'/2-j'}[-(1+i)\xi'] D_{i \lambda/2-j}[(1-i)\xi] \textrm{e}^{-i \omega t}\, .
\label{eq:nikishov-L}
\end{equation}
These integrals can be computed following the techniques of \cite{nikishov-71} (see Appendix). The exact result has the form:
\begin{align}
& L_{j'j} = \frac{\hbar}{A v_F} \exp\left( - i \omega \frac{p_1 + q_1}{2eE} \right) \textrm{e}^{-i \beta \varphi} \textrm{e}^{(j + j'-1)\varphi} K_{j'j}(\rho) \, ,  \nonumber \\
& K_{j'j}(\rho) = \int_{-\infty}^{+\infty} dv D_{-i\lambda'/2 - j'}[(1+i)v] D_{i\lambda/2-j}[-(1-i)v] \textrm{e}^{- i \rho v} \, ,
\label{eq:L-K}
\end{align}
with $\rho$ and $\varphi$ defined as in Eq.~\eqref{eq:rho-phi}, and
\begin{multline}
K_{j'j}(\rho) = \frac{\textrm{e}^{i \rho^2/4}}{\sqrt{2}} \Gamma(1 + i \beta -j-j') \, \text{e}^{\frac{\pi}{8}(\lambda-3\lambda')} \, 	\textrm{e}^{i \frac{\pi}{4}(j+3j')} (1 - \textrm{e}^{- \pi \lambda + \pi \lambda^\prime})\\ 
\times \left( \frac{-i \rho}{\sqrt{2}} \right)^{-1 -i \beta + j+j'} \Phi\left( i \frac{\lambda^\prime}{2} + j^\prime, -i\beta +j+j'; -i\frac{\rho^2}{2} \right)  \, .
\label{eq:K-integrals}
\end{multline}
The contiguity relations of confluent hypergeometric functions now lead to the identities (see proof in the Appendix):
\begin{align}
\textrm{e}^{i\pi/4} \frac{\rho}{\sqrt{2}} K_{00} & = - \frac{\lambda}{2} K_{01} - i \frac{\lambda'}{2} K_{10} \, , \nonumber \\
\textrm{e}^{i \pi/4} \frac{\rho}{\sqrt{2}} K_{11} & = - K_{01} - i K_{10} \, ,
\label{eq:contiguity-relations-K}
\end{align}
showing that only two of the $K_{j^\prime j}$ integrals are independent.

The identities \eqref{eq:contiguity-relations-K} allow the amplitudes $Q^{(pa)}_{\vect k \alpha}$ to be written in terms of $K_{01}$ and $K_{10}$ only. Substituting the expressions \eqref{eq:nikishov-L}--\eqref{eq:contiguity-relations-K} in the formula for $Q^{(pa)}_{\vect k \alpha}$ given in Eq.~\eqref{eq:amplitude-Q}, we obtain:
\begin{multline}
Q^{(pa)}_{\vect k \alpha} = \frac{\hbar}{A v_F} \textrm{e}^{i \beta \varphi} \exp\left(i \omega \frac{p_1+q_1}{2eE} \right) \left\{ K_{01}^\ast (1+i) \left[ A \chi_2 S_\alpha^{+-} + \frac{1}{\rho} \left( A^2 \lambda \textrm{e}^{-\varphi} + \pi_2^\ast \chi_2 \textrm{e}^{\varphi} \right) S_\alpha^{--}  \right] \right. \\
\left. + K_{10}^\ast (1-i)\left[ A \pi_2^\ast S_\alpha^{+-} + \frac{1}{\rho} \left( A^2 \lambda^\prime \textrm{e}^{-\varphi} + \pi_2^\ast \chi_2 \textrm{e}^{\varphi} \right) S_\alpha^{--} \right] \right\} \, .
\label{eq:explicit-Q-lambda}
\end{multline}
The corresponding differential transition probability is:
\begin{equation}
d\Gamma^{(pa)}_{\vect k \alpha} = \frac{e^2 v_F^2}{\hbar} \frac{2\pi}{V \omega}  (C C')^2 \left|Q^{(pa)}_{\vect k \alpha} \right|^2 \, S (2\pi)^2 \delta\left(\frac{\vec p}{\hbar} - \frac{\vec q}{\hbar} - \vec k\right) \frac{S d(\vec q/\hbar)}{(2\pi)^2} \, \frac{V d \vect k}{(2\pi)^3}\, .
\label{eq:transition-M-eh}
\end{equation}
For unpolarized emission, $\big|Q^{(pa)}_{\vect k \alpha} \big|^2$ should be replaced with
\begin{equation}
Q^2 = \sum_{\alpha=1}^2 \left|Q^{(pa)}_{\vect k \alpha} \right|^2 = h_0 |K_{01}|^2 + h_1 |K_{10}|^2 + 2 \, \textrm{Re}[h_2 K_{01}^\ast K_{10}] \, ,
\label{eq:Q-squared}
\end{equation}
where the functions $h_i$ are obtained from Eqs.~\eqref{eq:spin-S}, \eqref{eq:amplitude-Q}, \eqref{eq:L-K} and \eqref{eq:contiguity-relations-K},
\begin{equation}
h_0 = \frac{\lambda}{2}f_0 \, , \qquad h_1 = \frac{2}{\lambda} f_1 \, , \qquad h_2 = - f_2 \, .
\label{eq:h-coefficients}
\end{equation}
Integrating in $\vec q$-space, we find the unpolarized photon emission probability from an electron initially in the upper band undergoing an interband transition:
\begin{equation}
\frac{d\Gamma^{(pa)}}{d \Omega \, d\omega } = \frac{e^2}{\hbar c} \left(\frac{v_F}{c}\right)^2 \frac{\omega}{(2 \pi)^2}  (C C')^2 S^2 Q^2 \, .
\label{eq:transition-probability-eh}
\end{equation}

\subsection{Photon emission by electron in the lower band}

\subsubsection{Intraband transition}
\label{sec:lower-intra}

Consider the case of an electron with initial quasimomentum $\vec p$ and energy $E_p(t)<0$ in the asymptotic past, and final quasimomentum $\vec q$ and energy $E_q(t)<0$ in the asymptotic future. The amplitude of this process, to first order, is given by:
\begin{align}
M^{(h)}_{\vect k \alpha} &= \frac{1}{i\hbar} \int d\vect x \, dt \, \Bigl({} ^{-}\chi^\dagger_{\vec q} \otimes \bra{\vect k,\alpha} \Bigr) V_{int} \Bigl( {} _{-}\chi_{\vec p} \otimes \ket{0} \Bigr) \, .
\label{eq:h-amplitude-1}  \\
				&= i e \frac{v_F}{\hbar} \sqrt{\frac{2 \pi \hbar}{V \omega}} \left[ \int d\vec x \, \textrm{e}^{i(\vec p -\hbar \vec k - \vec q)\cdot \vec x/\hbar} \right] C C^\prime N^{(h)}_{\vect k \alpha} \, .
\label{eq:h-amplitude-2}
\end{align}
The coefficient $N^{(h)}_{\vect k \alpha}$ is the time integral of the time-dependent factors in \eqref{eq:h-amplitude-1}, as in previous cases. It can be shown that:
\begin{equation}
N^{(h)}_{\vect k \alpha} = \left[ N^{(e)}_{\vect k \alpha}(-\omega) \right]^\ast \, .
\label{eq:N-amplitude-lower}
\end{equation}
The amplitude $N^{(e)}_{\vect k \alpha}(-\omega)$ describes the absorption of a photon with wavenumber $-\vect k$ by an upper energy electron undergoing an intraband transition. Since the substitution $\omega \to -\omega$ also leads to the inversion of the signs of $\rho$ and $\varphi$ in the formulas for $N^{(e)}_{\vect k \alpha}$, the photon emission probability for the present case is obtained from Eq.~\eqref{eq:transition-probability-er} by changing the signs of $\rho$, $\varphi$ and $\omega$ in \eqref{eq:explicit-N-lambda}. The unpolarized photon emission is obtained in the same way from Eqs.~\eqref{eq:N-squared}--\eqref{eq:f10-full}. The transformation of the coefficients $f_i$ is trivially implemented, and the time integrals $I_{j'j'}$ satisfy:
\begin{equation}
I_{00}(-\rho) = \textrm{e}^{\pi \beta} I_{00}(\rho) \, , \qquad I_{11}(-\rho) = \textrm{e}^{\pi \beta} I_{11}(\rho) \, , \quad \text{for } \rho<0 \, .
\label{eq:I-rho-rev}
\end{equation}

\subsubsection{Interband transition}
\label{sec:lower-inter}

Finally, consider the case of an electron with initial quasimomentum $\vec p$ and energy $E_p(t)<0$ in the asymptotic past, and final quasimomentum $\vec q$ and energy $E_q(t)>0$ in the asymptotic future. The amplitude of the process, to first order, is given by:
\begin{align}
M^{(pc)}_{\vect k \alpha} &= \frac{1}{i\hbar} \int d\vect x \, dt \, \Bigl({} ^{+}\chi^\dagger_{\vec q} \otimes \bra{\vect k,\alpha} \Bigr) V_{int} \Bigl( {} _{-}\chi_{\vec p} \otimes \ket{0} \Bigr) \, .
\label{eq:pc-amplitude-1}  \\
				&= i e \frac{v_F}{\hbar} \sqrt{\frac{2 \pi \hbar}{V \omega}} \left[ \int d\vec x \, \textrm{e}^{i(\vec p -\hbar \vec k - \vec q)\cdot \vec x/\hbar} \right] C C^\prime Q^{(pc)}_{\vect k \alpha} \, .
\label{eq:pc-amplitude-2}
\end{align}
The coefficient $Q^{(pc)}_{\vect k \alpha}$ is the time integral of the time-dependent factors in \eqref{eq:h-amplitude-1}, and is given by:
\begin{equation}
Q^{(pc)}_{\vect k \alpha} = - \left[ Q^{(pa)}_{\vect k \alpha}(-\omega) \right]^\ast \, .
\label{eq:Q-amplitude-lower}
\end{equation}
The amplitude $Q^{(pa)}_{\vect k \alpha}(-\omega)$ describes the absorption of a photon with wavenumber $-\vect k$ by an electron in the upper energy band undergoing an interband transition. The differential transition probability is obtained from \eqref{eq:transition-M-eh} by changing the signs of $\rho$, $\varphi$ and $\omega$ in \eqref{eq:explicit-Q-lambda}. For the unpolarized photon emission, this prescription should be applied to Eqs.~\eqref{eq:Q-squared}--\eqref{eq:transition-probability-eh}. The time integrals $K_{j'j'}$ satisfy:
\begin{equation}
K_{01}(-\rho) = \textrm{e}^{-\pi \beta} K_{01}(\rho) \, , \qquad K_{10}(-\rho) = \textrm{e}^{-\pi \beta} K_{10}(\rho) \, , \quad \text{for } \rho<0 \, .
\label{eq:K-rho-rev}
\end{equation}


\section{Photon emission from a single particle}
\label{sec:single}

In this section we analyze the angular distribution of photon emission from a single electron interacting with the graphene lattice in the presence of a uniform electric field. This problem is solved by a direct application of the amplitudes derived in Section \ref{sec:electron-photon}, and we discuss their most relevant aspects in this simpler context before embarking on the analysis of the more intricate many-body problem in Section \ref{sec:many-body}. In particular, we discuss simplifications in the general solution in the case of massless particles, in the $v_F/c \ll 1$ approximation, and for particles moving parallel to the applied field. We also compare the angular distribution of the photon emission rate in a strong electric field with the free case. An analysis of the time development of the process of radiation formation is presented as a tool for the derivation of time-dependent photon emission rates from the amplitudes calculated in Section \ref{sec:electron-photon}.

\subsection{Total photon emission}
\label{sec:single-total}

The unpolarized photon emission probabilities for intra and interband transitions from an initial state ${} _{+}\chi_{\vec p}$ in the presence of a uniform electric field are given by Eqs.~\eqref{eq:transition-probability-er} and \eqref{eq:transition-probability-eh}, respectively. The total probability for photon emission is the sum of both contributions:
\begin{equation}
\frac{d\Gamma}{d \Omega \, d\omega} = \frac{e^2}{\hbar c} \left(\frac{v_F}{c}\right)^2 \frac{\omega}{(2 \pi)^2}  \frac{\textrm{e}^{-\pi (\lambda+\lambda')/4}}{4 A^4} (N^2+Q^2) \, ,
\label{eq:total-photon-emission}
\end{equation}
where, from \eqref{eq:N-squared}, \eqref{eq:Q-squared} and \eqref{eq:h-coefficients}, the squared amplitudes are:
\begin{equation}
N^2+Q^2 =  f_0 \left( |I_{00}|^2 + \frac{\lambda}{2}|K_{01}|^2 \right)+ f_1 \left( |I_{11}|^2 + \frac{2}{\lambda} |K_{10}|^2  \right) + 2 \mathrm{Re} [f_2 (I_{00}^\ast I_{11}-K_{01}^\ast K_{10})] \, .
\label{eq:squared-total}
\end{equation}
Let us consider some special cases of this formula in a more explicit manner. The coefficients $f_i$ appear in the calculation of differential transition probabilities of several first order processes in the presence of a uniform electric field, and the present analysis is also useful for other processes.

For massless particles, the coefficients $f_i$ become:
\begin{align}
f_0 & = 4  \left(\frac{A \hbar}{v_F}\right)^2 \left\{ 1 + \left(\frac{p_2 \textrm{e}^\varphi + q_2 \textrm{e}^{-\varphi}}{A \rho} \right)^2 - \sin^2 \theta_\gamma \left[ \sin \phi_\gamma + \cos \phi_\gamma \left(\frac{p_2 \textrm{e}^\varphi + q_2 \textrm{e}^{-\varphi}}{A \rho} \right) \right]^2 \right\} \, , 
\label{eq:f00-m0}
\\
f_1 & = p_2^2 q_2^2 \left(\frac{\hbar}{A v_F}\right)^2 \left\{ 1 + \left(\frac{p_2 \textrm{e}^{-\varphi} + q_2 \textrm{e}^\varphi}{A \rho} \right)^2 - \sin^2 \theta_\gamma \left[ \sin \phi_\gamma - \cos \phi_\gamma \left(\frac{p_2 \textrm{e}^{-\varphi} + q_2 \textrm{e}^\varphi}{A \rho} \right) \right]^2 \right\} \, , 
\label{eq:f11-m0}
\\
f_2 & = 2 i p_2 q_2 \left(\frac{\hbar}{v_F}\right)^2 \left\{ (1-\sin^2 \theta_\gamma \sin^2 \phi_\gamma) - \frac{1- \sin^2 \theta_\gamma \cos^2 \phi_\gamma}{(A\rho)^2} \left[ p_2 q_2 (\textrm{e}^{2\varphi} + \textrm{e}^{-2\varphi}) + p_2^2 + q_2^2 \right]  \right. \nonumber \\
& \hspace{85pt} \left. + \frac{1}{A\rho} \sin^2 \theta_\gamma \sin \phi_\gamma \cos \phi_\gamma (q_2 - p_2) \left(\textrm{e}^\varphi - \textrm{e}^{-\varphi} \right)  \right\} \, .
\label{eq:f01-m0}
\end{align}
The functions in curly brackets in \eqref{eq:f00-m0}--\eqref{eq:f01-m0} are all independent of $E$, since $A\rho$ and $\varphi$ are completely fixed by the wavenumber $\vect k$ of the emitted photon through
\begin{align}
A \, \rho \cosh \varphi & = - \hbar k \frac{c}{v_F} \, , \nonumber \\
A \, \rho \sinh \varphi & = \hbar k_1 \, ,
\label{eq:A-rho-k}
\end{align}
which follow from \eqref{eq:rho-phi} and quasimomentum conservation. The parameter $\varphi$, in particular, depends only on the direction of the radiation, $\tanh \varphi=-\sin \theta_\gamma \cos \phi_\gamma v_F/c$. So the $f_i$ depend on $E$ only through the powers of $A = (eE\hbar / v_F)^{1/2}$ outside the brackets. Moreover, there is no explicit dependence on the longitudinal components of the initial and final quasimomenta of the electron, which  appear only in the combination $p_1-q_1=\hbar k_1$. This allows one of the quasimomenta to be fixed arbitrarily with an appropriate choice of coordinates.

Since the Fermi velocity $v_F$ is much smaller than the speed of light $c$ ($c/v_F \simeq 300$), it is natural to consider an expansion in $v_F/c$. The hyperbolic angle $\varphi$ is always a small quantity, since $|\tanh \varphi| \leq v_F/c \simeq 1/300$. As a result, $\tanh \varphi \simeq \varphi$. The exponentials $\exp(\pm \varphi)$ can thus be set equal to $1$ to a good approximation, and the next terms in the power expansion of the exponential function may be added for further corrections of higher order in $v_F/c$. Moreover, according to \eqref{eq:A-rho-k}, the parameters $\rho$ and $\varphi$ can be approximated to first-order by:
\begin{align}
\frac{1}{A \rho} & \simeq - \frac{v_F}{c} \frac{1}{\hbar k} \, , \label{eq:low-v-approximation-rho} \\
\varphi & \simeq - \frac{v_F}{c} \sin \theta_\gamma \cos \phi_\gamma \, .
\label{eq:low-v-approximation-varphi}
\end{align}
Because of the explicit form of the formulas for the coefficients $f_i$, it is sufficient to apply the approximation \eqref{eq:low-v-approximation-rho} and set $\exp(\pm \varphi)\simeq 1$ in \eqref{eq:f00-m0}--\eqref{eq:f01-m0} in order to obtain an approximation valid to first-order in $v_F/c$.

A case of particular interest is that of massless particles with initial or final quasimomentum parallel to the applied field. In the asymptotic past or future, the transverse component of the physical momentum can be neglected, since the longitudinal component increases linearly with time. Therefore, for large times, the radiation from any state in the strong field regime $\lambda,\lambda' \ll 1$ can be approximated by that of a particle moving parallel to the field. A considerable simplification takes place in the formulas for the $f_i$ in this case. Put $p_2=0$, for instance. Then \eqref{eq:f11-m0} and \eqref{eq:f01-m0} give $f_1=f_2=0$. Moreover, from Eq.~\eqref{eq:I-integrals},
\begin{equation}
|I_{00}(\rho)|^2 = \pi \textrm{e}^{- 3 \pi \lambda /4 } \, ,
\label{eq:I00-zero-lambda}
\end{equation}
while, from Eq.~\eqref{eq:K-integrals},
\begin{equation}
|K_{01}(\rho)|^2 = \frac{2\pi}{\lambda} \textrm{e}^{-3\pi \lambda/4} \left(\textrm{e}^{\pi \lambda}-1\right) \, .
\label{eq:K01-zero-lambda}
\end{equation}
Inserting these values in \eqref{eq:total-photon-emission} and \eqref{eq:squared-total}, and noticing that for $p_2=0$,
\[
\frac{q_2 \textrm{e}^{-\varphi}}{A \rho} = - \frac{v_F}{c} \frac{\sin \theta_\gamma \sin \phi_\gamma}{1+\frac{v_F}{c} \sin \theta_\gamma \cos \phi_\gamma} \, ,
\]
we obtain a compact expression for the photon emission probability:
\begin{equation}
\frac{d\Gamma}{d \Omega \, d\omega} = \frac{e^2}{\hbar c} \left(\frac{v_F}{c}\right)^2 \frac{\omega}{4 \pi} \frac{\hbar}{eEv_F} \left\{ 1 - \left[ 1- \left( \frac{v_f}{c} \right)^2 \right] \frac{\sin^2 \theta_\gamma \sin^2 \phi_\gamma}{\left( 1+\frac{v_F}{c} \sin \theta_\gamma \cos \phi_\gamma \right)^2} \right\} \, .
\label{eq:total-emission-m0-lambda}
\end{equation}
This simple formula will be convenient for the comparison with the free case later. The contribution $N^2$ of the intraband process to the photon emission probability \eqref{eq:total-emission-m0-lambda} is proportional to $|I_{00}|^2$, and therefore exponentially attenuated for large $\lambda$, according to \eqref{eq:I00-zero-lambda}. The contribution $Q^2$ of the interband process is related in a simple manner to $N^2$:
\[
\frac{Q^2}{N^2} = \textrm{e}^{\pi \lambda}-1 \, .
\]
Hence, intraband transitions are the dominant mechanism of photon emission from a single particle in the strong field regime. For large $\lambda$, the situation reverses, and electron-hole annihilation (interband) become dominant. In the limiting case where the external electric field is absent, pair annihilation is the only process to be taken into account.

For an electron with initial state ${} _{-}\chi_{\vec p}$, the amplitude of photon emission is given by Eqs.~\eqref{eq:h-amplitude-2} and \eqref{eq:N-amplitude-lower} for intraband transitions, and by Eqs.~\eqref{eq:pc-amplitude-2} and \eqref{eq:Q-amplitude-lower} for interband transitions. The total probability for photon emission is the sum of the squared amplitudes of both contributions. According to the discussion in Sections \ref{sec:lower-intra} and \ref{sec:lower-inter}, the differential transition probability is obtained from \eqref{eq:total-photon-emission} and \eqref{eq:squared-total} through the substitutions $\rho \to -\rho$, $\varphi \to - \varphi$ and $\omega \to -\omega$ in the formulas for $|N|^2$ and $|Q|^2$. These can be applied directly in Eqs.~\eqref{eq:f00-m0}--\eqref{eq:f01-m0} to describe radiation from massless particles, and the approximation $v_F/c\ll 1$ can be implemented as before.

\subsection{Photon emission rate in a strong field}
\label{sec:single-rate}

The photon emission probability $d\Gamma/d\Omega$ given in Eq.~\eqref{eq:total-photon-emission} describes the probability $d\Gamma$ that a photon with frequency between $\omega$ and $\omega+d\omega$ is emitted in the solid angle $d\Omega$ by a conduction electron in graphene, regardless of the time at which the photon is emitted. If we restrict to a finite time interval $T=[t_1,t_2]$, however, only a fraction $d\Gamma(t_1,t_2)$ of the total emission will be observed. In the limit of an infinitesimal $\Delta t = t_2-t_1 \to dt$, we obtain the photon emission rate $d\Gamma/dt$. In this section, we compute $d\Gamma/dt$ for a conduction electron in the presence of a strong electric field using the results of the last section, and compare it to the free photon emission rate derived in \cite{mecklenburg}.

Because we treat the electric background nonperturbatively, the photon emission rate does not follow from an application of Fermi's golden rule as usual. The unperturbed states employed in the calculation of the amplitudes of photon emission are exact solutions of the Dirac equation in the presence of the electric field, having thus a nontrivial time-evolution, and as a result the energy of the particle is not conserved, invalidating Fermi's rule. An alternative strategy is thus required. We will present now an analysis of the time development of the process of radiation formation which can be used as a tool for the calculation of the emission rates. This technique will also be employed later in the derivation of the photon emission rate from the electronic gas at the Dirac point in the presence of a uniform electric field.

Before we start the calculations, let us outline the general procedure.  The contributions of intra and interband transitions to the photon emission probability $d\Gamma/d\Omega$ are described by the amplitudes $M_{\vect k \alpha}^{(e)}$ and $M_{\vect k \alpha}^{(pa)}$, which are defined by the time integrals \eqref{eq:M-e} and \eqref{eq:pa-amplitude-1}. These are defined over the whole real line, but each integral is actually dominated by a small interval of width $\Delta t_\omega$ around some $t_\omega$. We interpret $t_\omega \pm \Delta t_\omega$ as the time of formation of the corresponding radiation. It turns out that the radiation formation time depends only on $\omega$, as indicated by the notation. The time-dependent transition probability $d\Gamma(t_1,t_2)$ can then be defined as the squared amplitude of the superposition of those processes for which $t_\omega \in T$, i.e., those which actually occur in the interval under observation. Since the formation time $t_\omega$ is fixed by $\omega$, we are led in this way to a time-dependent cutoff in frequency space, which regulates the integration of $d\Gamma/d\Omega$. The photon emission rate is the limit of $d\Gamma(t_1,t_2)/(t_2-t_1)$ for small $(t_2-t_1) \to 0$.

We start the analysis with the case of intraband transitions. The amplitude $M^{(e)}_{\vect k \alpha}$ of this process is defined in \eqref{eq:M-e}, and requires the evaluation of the integral
\begin{equation}
\int d\vec x \, dt  {}\, ^{+}\psi^\dagger_{\vec q}  ( \vec \sigma \cdot \vec \epsilon_{\vect k \alpha} ) {} _{+}\psi_{\vec p} \, \textrm{e}^{-i (\vec k \cdot \vec x - \omega t)} \, .
\label{eq:J-time-formation}
\end{equation}
The in- and out-states in \eqref{eq:J-time-formation} have asymptotic behaviors:
\begin{align}
^{\pm}\psi_{\vec p} & \propto  \textrm{e}^{\mp i \xi'^2 /2} \, \textrm{e}^{i \vec p \cdot \vec x / \hbar} \, ,  & \text{for } t \gg \frac{p_1}{eE} + \sqrt{\frac{\hbar}{eEv_F}} (1+\lambda') \, , \nonumber \\
_{\pm}\psi_{\vec q} & \propto  \textrm{e}^{\pm i \xi^2 /2} \, \textrm{e}^{i \vec q \cdot \vec x / \hbar} \, ,  & \text{for } t \ll \frac{q_1}{eE} -\sqrt{\frac{\hbar}{eEv_F}} (1+\lambda) \, ,
\label{eq:asymptotic-approximations}
\end{align}
as can be checked from the asymptotic behavior of the Weber functions in \eqref{eq:asymptotic-states-dplus} and \eqref{eq:asymptotic-states-dminus} (see \cite{erdelyi}). Moreover, from \eqref{eq:g-matrices},
\begin{align}
_{+}\psi_{\vec p} =  g(^+ \mid _+)(\vec p) \, ^{+} \psi_{\vec p} + g(^- \mid _+)(\vec p) \, ^{-} \psi_{\vec p} \, , \nonumber \\
^{+}\psi_{\vec q} = g(_+ \mid ^+)(\vec q) \, _{+} \psi_{\vec q} + g(_- \mid ^+)(\vec q) \, _{-} \psi_{\vec q} \, .
\label{eq:bogolyubov-coefficients}
\end{align}
With an adequate choice of coordinates, we can set $p_1=0$, so we restrict to $\vec p = (0,p_2)$. At the end of the analysis, the longitudinal momentum can be easily reintroduced. We consider the cases $q_1< 0$ and $q_1>0$ independently.

Let $q_1<0$. The domain of the integral \eqref{eq:J-time-formation} can be decomposed into three convenient parts as $\mathbb{R}=(-\infty, q_1/eE) \cup [q_1/eE,0] \cup (0,\infty)$. In each of these regions, a simple approximation for the time-dependent part of the integrand follows from the exact transformation \eqref{eq:bogolyubov-coefficients} and the asymptotic approximations \eqref{eq:asymptotic-approximations}. In the region $(-\infty, q_1/eE)$, there are contributions of the form
\begin{equation}
g^\ast(_\pm \mid ^+)(\vec q) \int_{-\infty}^{q_1/eE} dt \, \textrm{e}^{i \xi^{\prime 2} /2} \textrm{e}^{\mp i \xi^2 /2} \textrm{e}^{i \omega t} \, .
\label{eq:time-bound}
\end{equation}
Up to a phase, the contribution proportional to $g^\ast(_- \mid ^+)$ corresponds to
\begin{equation}
g^\ast(_- \mid ^+) \int_{-\infty}^{q_1} dt \, \exp i \left[ \xi + \frac{1}{2} \left( \frac{q_1}{A} + \tau \right) \right]^2 \, ,
\label{eq:dominant-contribution}
\end{equation}
where $\tau = \omega A/eE$. The real and imaginary parts of \eqref{eq:dominant-contribution} are Fresnel integrals, which are dominated by a region of width $\Delta t_\omega \sim \sqrt{\hbar/eEv_F}$ around
\begin{equation}
\xi + \frac{1}{2} \left( \frac{q_1}{A} + \tau \right) \sim 0 \quad \implies \quad eE t_\omega \sim \frac{q_1}{2} - \frac{\hbar \omega}{2v_F} \, ,
\label{eq:stationary-phase}
\end{equation}
where the oscillations are slower. Now, from quasimomentum conservation, $q_1 = - \hbar k_1$. Since $\omega=k c$ and $c/v_F \gg 1$, it follows that $q_1/2$ is negligible, leading to the time of formation:
\begin{equation}
t^-_\omega \sim - \frac{\hbar \omega}{2eEv_F}  \, .
\label{eq:t-omega}
\end{equation}
The remaining contributions to the quantum amplitude can be analyzed in a similar manner, and are found to be negligible compared to \eqref{eq:dominant-contribution} (they do not contribute in the stationary phase approximation being considered). For $q_1>0$, we can decompose the $t$ axis into $\mathbb{R}=(-\infty, 0) \cup [0,q_1/eE] \cup (q_1/eE,\infty)$, and repeat the analysis, reaching the same result. We conclude that, for intraband transitions, radiation with frequency $\omega$ is emitted only at $t^{-}_\omega \pm \Delta t_\omega$. This contribution to the transition probability vanishes, however, if $|g(_- \mid ^+)(\vec q)|= 0$.

Consider now the case of interband transition. The amplitude $M^{(pa)}_{\vect k \alpha}$ of this process is defined in \eqref{eq:pa-amplitude-1}, which involves the integral:
\[
\int d\vec x \, dt  {}\, ^{-}\psi^\dagger_{\vec q}  ( \vec \sigma \cdot \vec \epsilon_{\vect k \alpha} ) {} _{+}\psi_{\vec p} \, \textrm{e}^{-i (\vec k \cdot \vec x - \omega t)} \, .
\]
We can repeat the analysis as before. For $q_1<0$, we have now integrals of the form
\[
g(_\pm \mid ^-)^\ast(\vec q) \int_{-\infty}^{q_1/eE} dt \, \textrm{e}^{i \xi^{\prime 2} /2} \textrm{e}^{\mp i \xi^2 /2} \textrm{e}^{i \omega t}
\]
in the region $(-\infty, q_1)$. The only difference with respect to the previous case is that the $g$ coefficients change, the integration to be performed being the same. The same arguments apply, and we find that the integral is dominated by an interval of width $\Delta t_\omega$ around $t^-_\omega$. This contribution vanishes if $|g(_- \mid ^-)(\vec q)| = 0$. But now we also have a contribution proportional to $g(^+ \mid _+)(\vec p)$ from the interval $(0,\infty)$, corresponding to the emission of radiation of frequency $\omega$ at the instant
\begin{equation}
t^+_\omega \sim \frac{\hbar \omega}{2eEv_F}  \, .
\label{eq:t-omega-eh}
\end{equation}
The same results are obtained for $q_1>0$. We conclude that for interband transitions, radiation with frequency $\omega$ is emitted at the intervals $t^{\pm}_\omega \pm \Delta t_\omega$.

For $p_1 \neq 0$, the radiation formation is simply translated by $p_1/eE$, leading to the formation times
\begin{equation}
t^{\pm}_\omega \simeq \frac{p_1}{eE} \pm \frac{\hbar \omega}{2eEv_F} \, ,
\label{eq:radiation-time}
\end{equation}
around which radiation with frequency $\omega$ is produced in an interval of width 
\begin{equation}
\Delta t_\omega = \sqrt{\hbar/eEv_F} \, .
\label{eq:formation-width}
\end{equation}
For a strong electric field $E$, the width of the formation time is narrow, and the spectrum of the radiation emitted in a finite interval of time is restricted by \eqref{eq:radiation-time}. If $E \sim 10^6 \text{ V/m}$, for instance, $\Delta t_\omega \sim 10^{-14} \text{ s}$. In a vanishingly weak field, the formation time becomes infinite, and the above analysis does not apply.

Now let us apply Eq.~\eqref{eq:radiation-time} to the derivation of the photon emission rate $d\Gamma/dt$ from a massless conduction electron with initial quasimomentum $\vec p$ in the presence of a strong electric field. We take $\vec p = (p_1,0)$, and let $p_1>0$. We are interested in comparing the angular distribution of the radiation in a strong field with the free case \cite{mecklenburg}, so we restrict to $\lambda\ll 1$.  In this regime, $|g(_- \mid ^-)(\vec q)|$ and $|g(^+ \mid _+)(\vec p)| \simeq 0$, and photon emission is dominated by intraband transitions. Radiation of frequency $\omega$ is formed at $t_\omega^-$, according to the analysis of radiation formation. From \eqref{eq:radiation-time}, the spectrum of the radiation emitted during $T=[t_1,t_2]$ is restricted to the range of frequencies $F=[\omega_2,\omega_1]$, with
\begin{equation}
\omega_i = 2\frac{v_F}{\hbar} (p_1- eEt_i)  \, , \quad i=1,2 \, .
\label{eq:time-cutoff}
\end{equation}
Hence, the integration of $d\Gamma/d\Omega\,d\omega$ should be restricted to the interval $F$ with width $\Delta \omega = (2eEv_F /\hbar) \Delta t$. Using Eq.~\eqref{eq:total-emission-m0-lambda}, we obtain:
\begin{equation}
\left. \frac{d\Gamma}{d \Omega}(t_1,t_2)\right|_{m=0,p_2=0} = \frac{e^2}{\hbar c} \left(\frac{v_F}{c}\right)^2 \frac{\hbar}{eEv_F} \left\{ 1 - \left[ 1- \left( \frac{v_f}{c} \right)^2 \right] \frac{\sin^2 \theta_\gamma \sin^2 \phi_\gamma}{\left( 1+\frac{v_F}{c} \sin \theta_\gamma \cos \phi_\gamma \right)^2} \right\} \frac{(\omega_1^2-\omega_2^2)}{8\pi} \, .
\label{eq:T-strong-parallel}
\end{equation}
For a small $\Delta t$ satisfying
\begin{equation}
eE \Delta t \ll |p_1 - eEt_0| \, , \qquad t_0 = \frac{t_1 + t_2}{2} \, ,
\label{eq:rate-condition}
\end{equation}
the photon frequency $\omega_0$ has negligible variation in $T$, and all radiation is emitted approximately at the same frequency
\[
\omega \simeq \omega_0 = 2 \frac{v_f}{\hbar} |p_1 - e E t_0| \, , \qquad \text{for } t \in [t_1,t_2] \, .
\]
In this case, we obtain from \eqref{eq:T-strong-parallel} the strong field photon emission rate:
\begin{equation}
\left. \frac{d\Gamma}{d \Omega dt}\right|_{m=0,p_2=0} = \frac{e^2}{\hbar c} \left(\frac{v_F}{c}\right)^2 \frac{\omega_0}{2 \pi} \left\{ 1 - \left[ 1- \left( \frac{v_f}{c} \right)^2 \right] \frac{\sin^2 \theta_\gamma \sin^2 \phi_\gamma}{\left( 1+\frac{v_F}{c} \sin \theta_\gamma \cos \phi_\gamma \right)^2} \right\} \, .
\label{eq:rate-strong-parallel}
\end{equation}
This formula can be compared with the free photon emission rate derived in \cite{mecklenburg}.

To the lowest order in $v_F/c$, the angular dependence in \eqref{eq:rate-strong-parallel} has the simple form $1-\sin^2 \theta_\gamma \sin^2 \phi_\gamma$. At this level of approximation, the photon emission rate \eqref{eq:rate-strong-parallel} is identical to that obtained in the free case in \cite{mecklenburg} at the same approximation. (In order to perform the comparison, write Eq.~(13) of \cite{mecklenburg} in terms of the photon frequency using energy conservation, and set $\phi_c=-\pi$, since the physical momentum $p^1 +eEt^-_\omega$ is negative at $t^-_\omega$.) This shows that the photon emission rate is not affected by the electric field to the leading order in $v_F/c$, even in the presence of a strong field. The first-order correction in $v_F/c$ is distinct in the two cases, however. In the strong field regime, it is of the form $2 (v_F/c) \sin^3 \theta_\gamma \sin^2 \phi_\gamma \cos \phi_\gamma$, while in the free case it is of the form $2 (v_F/c) \sin \theta_\gamma \cos \phi_\gamma$. Therefore, if the angular distribution of photon number count is measured with a precision of the order $v_F/c \simeq 0.3\%$, it is necessary to take into account the effects of the electric field, and Eq.~\eqref{eq:rate-strong-parallel} should be used.

\section{Photon emission at the Dirac point}
\label{sec:many-body}

At zero temperature and chemical potential, the Fermi level of pristine graphene is at the charge neutrality Dirac point. Let a uniform electric field parallel to the plane of the material be switched on at $t=0$, and act for a duration of time $T$. Electron-hole pairs are then created by the applied field, and these charged excitations emit radiation. In this section we describe such photon emission induced by the electric field. As in previous sections, the effect of the electric background is taken into account exactly, and the interaction with the quantized electromagnetic field is considered to first-order in perturbation theory. For technical matters, we first study the photon emission from many-particle states constructed from exact in- and out-solutions of the Dirac equation in a constant electric field, and then introduce appropriate time-dependent cutoffs which lead to the desired result for a field of finite duration.

\subsection{Many-particle states and first-order processes}

Two complete sets of exact solutions of the Dirac equation in a constant electric field were introduced in Eqs.~\eqref{eq:asymptotic-states-dplus}--\eqref{eq:asymptotic-3d} in Section \ref{sec:settings}. Let $_{\zeta}a^\dagger_{\vec p} \, , {} _{\zeta}a_{\vec p}$ be creation and annihilation operators associated with the in-solutions $_{\zeta}\chi_{\vec p}(t, \vect x)$, and $^{\zeta}a^\dagger_{\vec p} \, , {} ^{\zeta}a_{\vec p}$ be creation and annihilation operators associated with the out-solutions $^{\zeta}\chi_{\vec p}(t, \vect x)$. These operators allow us to construct two distinct Fock representations $\mathcal{H}_{in}$ and $\mathcal{H}_{out}$ of the Hilbert space of many-electron states in graphene, which we call the in- and out-representations. In order to do so, we introduce in- and out-vacua
\begin{align}
\ket{0,in} &= \prod_{\vec s} {} _{-}a^\dagger_{\vec s} \, \ket{bare} \, , 
\label{eq:in-vacuum} \\ 
\ket{0,out} &= \prod_{\vec s} {} ^{-}a^\dagger_{\vec s} \, \ket{bare}  \, ,
\label{eq:out-vacuum}
\end{align}
where $\ket{bare}$ is the state in which there are no free electrons in the honeycomb lattice, and apply the corresponding creation and annihilation operators to represent quasiparticle excitations of each vacua. An arbitrary state in $\mathcal{H}_{in}$ is a finite-norm superposition of quasiparticle excitations of the in-vacuum, and similarly for $\mathcal{H}_{out}$. States of the quantized electromagnetic field are represented as before.

In the in-vacuum $\ket{0,in}$, all single-particle in-states ${} _{-}\chi_{\vec p}$ with asymptotically negative energy are occupied. Hence, for any given $\vec p$, the state with instantaneous momentum $P_i(t)=p_i- eEt \, \delta_{i1}$ in the lower energy band is occupied for sufficiently large negative times $t<p_1/eE$, while the corresponding state in the upper energy band is free. In this sense, $\ket{0,in}$ approaches the ground state of the electronic gas in graphene in the asymptotic past, allowing it to be used under certain conditions for the calculation of processes with initial state at the charge neutrality point at some finite time $t_0$. The convergence is not uniform, however, and the approximation should be restricted to processes involving only quasimomenta $\vec p$ such that $p_1>eEt_0$, as we will discuss later.

The many-body interaction Hamiltonian describing the coupling of the electron gas with the quantized electromagnetic field can be written as (see \cite{martin}, for instance):
\begin{equation}
\tilde{V}_{int}(t) = \sum_{\vec q,\zeta;\vec p, \zeta '} \left( \int d\vect x \; {} ^{\zeta}\chi^\dagger_{\vec q} \, V_{int} \; {} _{\zeta '}\chi_{\vec p} \right) {}^{\zeta}a^\dagger_{\vec q} \; {}_{\zeta '} a_{\vec p} \, ,
\label{eq:many-particle-H}
\end{equation}
where $V_{int}$ is the single-particle interaction Hamiltonian defined in Eq.~\eqref{eq:perturbation}. We have chosen a mixed representation in which the matrix elements of $V_{int}$ are computed between states in the in-representation (at the right) and in the out-representation (at the left). The interaction $\tilde{V}_{int}$ preserves the number of electrons, but can produce or annihilate photons, since $V_{int}$ couples single-electron states with the quantized electromagnetic field, and can also produce electron-hole quasiparticle excitations.

To first-order in time-dependent perturbation theory, an initial state $\ket{i}$ at the asymptotic past evolves in the asymptotic future into
\begin{equation}
U^{(1)}\ket{i} = \ket{i} + \frac{1}{i\hbar} \int_{-\infty}^{+\infty} dt \, \tilde{V}_{int} \ket{i} \, ,
\label{eq:1st-order-evolution-gen}
\end{equation}
where $U^{(1)}$ is the first-order unitary evolution operator associated with $\tilde{V}_{int}$. Such time evolution can be completely described in terms of the amplitudes of single-particle processes studied in Section \ref{sec:electron-photon} and the $g$ matrices defined in Eq.~\eqref{eq:g-matrices}, as we will show now.

A generic initial state $\Psi$ of the many-electron system can be represented in $\mathcal{H}_{in}$ in a linear basis formed by excitations of the in-vacuum of the form
\begin{equation}
\ket{{\vec r_{(1)}^{\;+}, \cdots , \vec r_{(K)}^{\;+}}, \vec s_{(1)}^{\;-}, \cdots ,\vec s_{(L)}^{\;-}\, ; \, in}={}_+ a^\dagger_{\vec r_{(1)}} \dots {}_+ a^\dagger_{\vec r_{(K)}} {}_- a_{\vec s_{(1)}} \dots {}_- a_{\vec s_{(L)}} \ket{0,in} \, .
\label{eq:many-in-state}
\end{equation}
The state \eqref{eq:many-in-state} corresponds to a configuration in which there are $K$ occupied modes in the upper energy band and $L$ holes in the Dirac sea in the asymptotic past. To zeroth-order, time evolution is dictated by the applied electric field. In order to describe it, we use the (Bogoliubov) transformation
\begin{equation}
{}_\zeta a^\dagger_{\vec p} = g(^+ \mid _\zeta) \, {}^+ a^\dagger_{\vec p} + g(^- \mid _\zeta) \, {}^- a^\dagger_{\vec p} \, ,
\label{eq:bogoliubov-io}
\end{equation}
which follows from \eqref{eq:g-matrices}, to map states in the in-representation $\mathcal{H}_{in}$ into states in the out-representation $\mathcal{H}_{out}$. In particular,
\begin{align}
\ket{0,in} &= \prod_{\vec s} {} _{-}a^\dagger_{\vec s} \, \ket{bare} 
\notag \\
&= \prod_{\vec s} \left[ g(^+ \mid _-)(\vec s) \, {}^+ a^\dagger_{\vec s} \, {}^- a_{\vec s} + g(^- \mid _-)(\vec s) \right] \ket{0,out} \, .
\label{eq:0th-order-many}
\end{align}
For each $\vec s$, the g-coefficients $g(^+ \mid _-)$ and $g(^- \mid _-)$ describe occupation numbers of the upper and lower energy states in the asymptotic future, respectively. Since in the asymptotic past only negative energy states are occupied, the first term in \eqref{eq:0th-order-many} describes the amplitude of electron-hole creation for each mode. The corresponding probability is $|g(^+ \mid _-)(\vec s)|^2=\textrm{e}^{-\pi \lambda}$, as expected. The second term in \eqref{eq:0th-order-many} describes the amplitude of the persistence of the mode $\vec s$ in the lower band, with probability $1-\textrm{e}^{-\pi \lambda}$. Applying the Bogoliubov transformation \eqref{eq:bogoliubov-io} to the creation and annihilation operators acting on the in-vacuum in \eqref{eq:many-in-state}, and taking Eq.~\eqref{eq:0th-order-many} into account, we can map any in-state to the out-representation.

Applying the first-order evolution operator \eqref{eq:1st-order-evolution-gen} to an initial state 
\[
\ket{i} = \ket{{\vec r_{(1)}^{\;+}, \cdots , \vec r_{(K)}^{\;+}}, \vec s_{(1)}^{\;-}, \cdots ,\vec s_{(L)}^{\;-}\, ; \, in} \otimes \ket{0}_{ph}
\]
with no initial photons, we obtain:
\begin{equation}
U^{(1)}\ket{i} = \ket{i} + \sum_{\vect k \alpha} \sum_{\vec q,\zeta;\vec p, \zeta '} M_{\vect k \alpha}(\vec p,\zeta ';\vec q, \zeta) \, {}^{\zeta}a^\dagger_{\vec q} \; {}_{\zeta '} a_{\vec p} \, c^\dagger_{\vect k \alpha} \, \ket{i} \, ,
\label{eq:1st-order-many-in}
\end{equation}
where
\begin{align}
M_{\vect k \alpha}(\vec p,+;\vec q,+) &= M^{(e)}_{\vect k \alpha}(\vec p,\vec q) \, , \notag \\
M_{\vect k \alpha}(\vec p,+;\vec q,-) &= M^{(pa)}_{\vect k \alpha}(\vec p,\vec q) \, , \notag \\
M_{\vect k \alpha}(\vec p,-;\vec q,+) &= M^{(pc)}_{\vect k \alpha}(\vec p,\vec q) \, , \notag \\
M_{\vect k \alpha}(\vec p,-;\vec q,-) &= M^{(h)}_{\vect k \alpha}(\vec p,\vec q) \, .
\label{eq:many-to-single}
\end{align}
Furthermore, the expression
\[
{}_{\zeta '} a_{\vec p} \, c^\dagger_{\vect k \alpha} \, \ket{i} =  {}_{\zeta '} a_{\vec p} \, {}_+ a^\dagger_{\vec r_{(1)}} \dots {}_+ a^\dagger_{\vec r_{(K)}} {}_- a_{\vec s_{(1)}} \dots {}_- a_{\vec s_{(L)}} \ket{0,in} \otimes \ket{\vect k \alpha}
\]
appearing in each term of the sum in Eq.~\eqref{eq:1st-order-many-in} can be transformed into the out-representation using Eqs.~\eqref{eq:bogoliubov-io} and \eqref{eq:0th-order-many}, leading to a formula for $U^{(1)}\ket{i}$ in the out-representation with linear coefficients written in terms of the single-particle amplitudes \eqref{eq:many-to-single} and the $g$-coefficients. If the initial state $\ket{i}$ has only a few excitations, the calculation of $U^{(1)}\ket{i}$ can be easily carried out. For more complex states, the combinatorics of creation and annihilation operators mixed with the Bogoliubov transformations \eqref{eq:bogoliubov-io} may become exceedingly complicated.

If one is interested only in the radiation emitted by the electrons, however, regardless of the final state of the many-electron system, it is not necessary to transform the final state $U^{(1)}\ket{i}$ to the out-representation. In this case one can just project such state as given in \eqref{eq:1st-order-many-in} into the one-photon subspace associated with $\ket{\vect k \alpha}$ and calculate the probabilities $|\bra{\vect k \alpha} U^{(1)}\ket{i}|^2$. These can be calculated in the in-representation. It is sufficient to apply the Bogoliubov transformation
\begin{equation}
{}^\zeta a^\dagger_{\vec p} = g(_+ \mid ^\zeta) \, {}_+ a^\dagger_{\vec p} + g(_- \mid ^\zeta) \, {}_- a^\dagger_{\vec p} \, ,
\label{eq:bogoliubov-oi}
\end{equation}
in the sum appearing in \eqref{eq:1st-order-many-in}, and then reduce the expression
\begin{multline*}
\bra{\vect k \alpha}U^{(1)}\ket{i} = \sum_{\vec q,\zeta;\vec p, \zeta '} M_{\vect k \alpha}(\vec p,\zeta ';\vec q, \zeta) \, \left[ g(_+ \mid ^\zeta) \, {}_+ a^\dagger_{\vec p} + g(_- \mid ^\zeta) {}_- a^\dagger_{\vec p} \right] \; {}_{\zeta '} a_{\vec p} \\
\times 
\, {}_+ a^\dagger_{\vec r_{(1)}} \dots {}_+ a^\dagger_{\vec r_{(K)}} {}_- a_{\vec s_{(1)}} \dots {}_- a_{\vec s_{(L)}} \ket{0,in}
\end{multline*}
into a sum of linearly independent terms using the standard anti-commutation relations among creation and annihilation in-operators. We will illustrate this procedure in the calculation of photon emission from the in-vacuum in the next section.

\subsection{Photon emission from in-vacuum}

Let the initial state in the asymptotic past be
\begin{equation}
\ket{i}=\ket{0,in} \otimes \ket{0}_{ph} \, .
\label{eq:initial-state}
\end{equation}
Applying the first-order time-evolution operator \eqref{eq:1st-order-evolution-gen} to such state, we obtain
\begin{equation}
U^{(1)}\ket{i} = \ket{i} + \sum_{\vec p, \vec q, \vect k, \alpha} \left[ M^{(h)}_{\vect k \alpha} \; {} ^{-}a^\dagger_{\vec q} \, {}_{-}a_{\vec p} \, + M^{(pc)}_{\vect k \alpha} \; {} ^{+}a^\dagger_{\vec q} \; {}_{-}a_{\vec p} \, \right] \ket{0,in} \otimes \ket{\vect k,\alpha} \, ,
\label{eq:1st-order-evolution}
\end{equation}
where $M^{(h)}_{\vect k \alpha}$ and $M^{(pc)}_{\vect k \alpha}$ are the amplitudes calculated in Section \ref{sec:electron-photon}. The summation in \eqref{eq:1st-order-evolution} corresponds to a superposition of one-photon states created by the interaction with the quantized electromagnetic field accompanied by excitations of the original in-vacuum induced by the electric background. The probability density that a photon $\ket{\vect k \alpha}$ is emitted, regardless of the number of electron-hole pairs produced, is given by the squared amplitude
\begin{equation}
d\Gamma^{in}_{\vect k \alpha} =  \left| \bra{\vect k \alpha} U^{(1)}\ket{i} \right|^2 \, \frac{V d \vect k}{(2\pi)^3} \, .
\label{eq:total-DP-def}
\end{equation}
For any $\vec k \neq 0$ we can use the identities
\begin{align}
{} ^{-}a^\dagger_{\vec q} \; {}_{-}a_{\vec p} \, \ket{0,in} = g(_+ \mid^-)(\vec q) \, {} _{+}a^\dagger_{\vec q} \, {}_{-}a_{\vec p} \ket{0,in} \, , \\
{} ^{+}a^\dagger_{\vec q} \, {}_{-}a_{\vec p} \, \ket{0,in} = g(_+ \mid^+)(\vec q) \; {} _{+}a^\dagger_{\vec q} \, {}_{-}a_{\vec p} \ket{0,in} \, , 
\end{align}
together with \eqref{eq:h-amplitude-1} and \eqref{eq:pc-amplitude-1}, in order to write the projection of the final state in the subspace with one photon $\ket{\vect k \alpha}$ in the form
\begin{equation}
\bra{\vect k \alpha} U^{(1)}\ket{i} = \sum_{\vec p, \vec q} M^{in}_{\vect k \alpha} (\vec p, \vec q)  \, {} _{+}a^\dagger_{\vec q} \, {}_{-}a_{\vec p} \ket{0,in}
\label{eq:partial-projection}
\end{equation}
with
\begin{align}
M^{in}_{\vect k \alpha} (\vec p, \vec q) &= \frac{1}{i \hbar} \int dt \, \Bigl({} _{+}\chi^\dagger_{\vec q} \otimes \bra{\vect k,\alpha} \Bigr) V_{int} \Bigl( {} _{-}\chi_{\vec p} \otimes \ket{0} \Bigr) \notag \\
&= i e \frac{v_F}{\hbar} \sqrt{\frac{2 \pi \hbar}{V \omega}} \int d\vec x \, dt  {}\, _{+}\psi^\dagger_{\vec q}  ( \vec \sigma \cdot \vec \epsilon_{\vect k \alpha} ) {} _{-}\psi_{\vec p} \, \textrm{e}^{-i (\vec k \cdot \vec x - \omega t)}.
\label{eq:amplitude-DP-1}
\end{align}

The amplitude $M^{in}_{\vect k \alpha}$ can be computed as the amplitudes of photon emission from single-particle states in Section \ref{sec:electron-photon}. We first write it as:
\begin{equation}
M^{in}_{\vect k \alpha} (\vec p, \vec q) = i e \frac{v_F}{\hbar} \sqrt{\frac{2 \pi \hbar}{V \omega}} \left[ \int d\vec x \, \textrm{e}^{i(\vec p -\hbar \vec k - \vec q)\cdot \vec x/\hbar} \right] C C^\prime N^{in}_{\vect k \alpha}(\vec p, \vec q) \, ,
\label{amplitude-DP-2}
\end{equation}
where $N^{in}_{\vect k \alpha}(\vec p, \vec q)$ is the integral of the time-dependent factors in \eqref{eq:amplitude-DP-1},
\begin{multline}
N^{in}_{\vect k \alpha}(\vec p, \vec q) =  2i A^2 S_{00}(-\omega) \, S_\alpha^{-+} + (1-i) A \pi_2 S_{10}(-\omega) \, S_\alpha^{--}
\\
- (1-i) A \chi_2 S_{01}(-\omega) S_\alpha^ {++} + \pi_2 \chi_2 S_{11}(-\omega) S_{\alpha}^{+-} \, ,
\label{eq:amplitude-DP-S}
\end{multline}
and the $S$ integrals are defined by:
\begin{equation}
S_{j'j}(\omega) = \int_{-\infty}^{+ \infty} dt \, D_{-i \lambda'/2-j'}[-(1+i)\xi'] D_{-i \lambda/2-j}[-(1+i)\xi] \textrm{e}^{-i \omega t}\, .
\label{eq:nikishov-s}
\end{equation}
Then we factor out the dependence on the longitudinal momenta $p_1,q_1$ (see Appendix),
\begin{equation}
S_{j'j}(\omega) = \frac{\hbar}{A v_F} \exp\left(- i \omega \frac{p_1+q_1}{2eE} \right) \textrm{e}^{(j'-j)\varphi} \textrm{e}^{-i\beta \varphi} R_{j'j}(\rho) \, ,
\label{eq:S-R}
\end{equation}
where $\rho,\varphi$ are defined in \eqref{eq:rho-phi}. Explicit formulas for the $R_{j'j'}(\rho)$ can be written in terms of confluent hypergeometric functions,
\begin{multline}
R_{j'j}(\rho) = \sqrt{\pi} \textrm{e}^{-i \pi/4} \textrm{e}^{i \rho^2/4} \\
\times \left[ \frac{\Gamma(-i\beta-j+j')}{\Gamma \left(i\frac{\lambda'}{2}+ j'\right)} \textrm{e}^{-\frac{\pi}{8}(\lambda'+3\lambda)} \textrm{e}^{i \frac{\pi}{4}(j'+3j)} \left(\frac{- i \rho}{\sqrt{2}}\right)^{i\beta+j-j'} \Phi\left(i \frac{\lambda}{2}+j, 1+ i\beta+j-j';-i \frac{\rho^2}{2} \right) \right. \\
\left.	+ \frac{\Gamma(i\beta+j-j')}{\Gamma \left(i\frac{\lambda}{2}+j \right)} \textrm{e}^{-\frac{\pi}{8}(\lambda+3\lambda')} \textrm{e}^{i \frac{\pi}{4}(j+3j')}  \left(\frac{- i \rho}{\sqrt{2}}\right)^{-i\beta-j+j'} \Phi\left(i \frac{\lambda'}{2}+j', 1- i\beta-j+j';-i \frac{\rho^2}{2} \right) \right] \, .
\label{eq:R-integrals}
\end{multline}
The $R_{j'j}$ integrals satisfy the contiguity relations
\begin{align}
R_{01}(-\rho) &= \frac{1-i}{\rho} R_{00}(-\rho) + \frac{1-i}{\rho} \left( \frac{i\lambda'}{2} \right) R_{11}(-\rho) \, , \notag \\
R_{10}(-\rho) &= \frac{1-i}{\rho} R_{00}(-\rho) + \frac{1-i}{\rho} \left( \frac{i\lambda}{2} \right) R_{11}(-\rho) \, ,
\label{eq:R-contiguity}
\end{align}
which allow us to write $N^{in}_{\vect k \alpha}$ in terms of $R_{00}$ and $R_{11}$ only,
\begin{multline}
N^{in}_{\vect k \alpha}(\vec p, \vec q) = \frac{\hbar}{A v_F} \textrm{e}^{i \beta \varphi} \exp\left(i \omega \frac{p_1+q_1}{2eE} \right)  \left\{ R_{00}(-\omega) \left[ 2 i A^2 S_\alpha^{-+} + 2i\frac{A}{\rho} \left( \chi_2 \textrm{e}^{\varphi} + \pi_2 \textrm{e}^{-\varphi} \right) S_\alpha^{++}  \right] \right. 
\\
\left. + R_{11}(-\omega) \left[ \pi_2 \chi_2 S_\alpha^{+-} - \frac{A}{\rho} \left( \lambda' \chi_2 \textrm{e}^{\varphi} + \lambda \pi_2 \textrm{e}^{-\varphi} \right) S_\alpha^{++} \right] \right\} \, .
\label{eq:amplitude-DP-R}
\end{multline}

Substituting \eqref{eq:partial-projection} and \eqref{amplitude-DP-2} in the formula \eqref{eq:total-DP-def} for the probability density of photon emission, and summing over polarizations, we obtain in the large area limit:
\begin{equation}
\frac{1}{S} \frac{d\Gamma^{in}}{d\Omega \, d\omega} =  \frac{e^2}{\hbar c} \left( \frac{v_F}{c} \right)^2 \frac{\omega}{(2\pi)^2} \frac{1}{4 A^4} \int \,  \frac{d\vec p}{(2\pi \hbar)^2} \, \textrm{e}^{-\pi (\lambda+\lambda')/4} O^2(\vec p, \vec q) \, ,
\label{eq:Gamma-DP}
\end{equation}
where the delta of quasimomentum conservation was used to eliminate the integration in $\vec q$, and
\begin{equation}
O^2(\vec p, \vec q) = \sum_{\alpha=1}^2 \left| N^{in}_{\vect k \alpha}(\vec p, \vec q) \right|^2 = \tilde{f}_0 |R_{00}(-\rho)|^2 + \tilde{f}_1 |R_{11}(-\rho)|^2 + 2 \mathrm{Re} [\tilde{f}_2 R_{00}^\ast(-\rho) R_{11}(-\rho)] \, ,
\label{eq:squared-unpol-DP}
\end{equation}
with $\tilde{f}_i = f_i(-\rho,-\varphi)$, and the $f_i$ are given by Eqs.~\eqref{eq:f00-full}--\eqref{eq:f10-full}. This is our general formula for the photon emission probability from the initial state $\ket{0,in}\otimes\ket{0}_{ph}$. Simplifications in the formulas for the coefficients $f_i$ in the case of massless particles, in the low $v_F/c$ approximation, and for longitudinal quasimomenta were discussed in Section \ref{sec:single-total}, and also apply to the present case.

\subsection{Initial state at Dirac point}

Let us study now the problem of a uniform electric field of finite duration $T$ switched on at $t=0$. At zero temperature and chemical potential, the ground state of the electronic gas in graphene is at the Dirac point for $t<0$, with the lower energy band completely filled and the upper energy band completely free. After the electric field is switched on, electrons begin to tunnel to the upper band, giving rise to charged excitations which are accelerated by the applied field and may annihilate, thereby emitting radiation. Such photon emission can be described using the formulas \eqref{eq:Gamma-DP} and \eqref{eq:squared-unpol-DP} for the photon emission probability from an initial state $\ket{0,in}$ in a constant uniform electric field with the introduction of appropriate cutoffs in Eq.~\eqref{eq:Gamma-DP}, as we will show now.

Exact solutions of the Dirac equation in the presence of a uniform electric field acting for a finite duration of time $T$ were studied in \cite{GG95}. In that work, the system was treated as a transmission problem, in which free states for $t<0$ and $t>T$ are connected by exact solutions of the Dirac equation in a uniform electric field in the region $\mathcal{T}=[0,T]$. Quasimomentum is conserved along the evolution, and a state initially in the lower (or upper) band ends up in a superposition of positive and negative states with the same quasimomentum. Band transition takes place essentially at $t_{bt}=p_1/eE$ \cite{nikishov-71,GG95}, and is restricted to quasimomenta $\vec p$ satisfying the condition:
\begin{equation}
0 < p_1 < eET \, .
\label{eq:finite-duration-cutoff}
\end{equation}
Accordingly, for an initial many-particle state at the Dirac point at $t=0$, there is no electron-hole creation for quasimomenta such that $p_1 <0$ or $p_1>eET$. We assume that these modes do not contribute significantly to photon emission. In the pair creation region selected by \eqref{eq:finite-duration-cutoff}, solutions of the Dirac equation with quasimomentum $\vec p$ and negative initial energy are well approximated, for $t \in [0,T]$, by the in-solutions ${} _{-}\chi_{\vec p} \,$ obtained for a constant electric field. Since these are the modes responsible for photon emission, it follows that the radiation emitted while the finite duration field is on corresponds to that emitted in the same interval of time by the modes satisfying \eqref{eq:finite-duration-cutoff} for an initial state in the in-vacuum $\ket{0,in}$ in the presence of a constant electric field. In short, the cutoff \eqref{eq:finite-duration-cutoff} encodes the condition that pair creation should take place in order that a particular mode contributes significantly to photon emission; for such modes, $\ket{0,in}$ is a good approximation to the ground state at $t=0$.

A second cutoff follows from an analysis of radiation formation. Let $\vec p$ be in the pair-creation region \eqref{eq:finite-duration-cutoff}. The contribution of this mode to the probability of photon emission \eqref{eq:Gamma-DP} is given by the square of the amplitude $M^{in}_{\vect k \alpha}$ introduced in \eqref{eq:amplitude-DP-1}. As in the case of a single electron in the graphene lattice, the probability of the emission of a photon with frequency $\omega$ is considerable only in a short interval of time $t^+_\omega \pm \Delta t_\omega$, which dominates the time integral in \eqref{eq:amplitude-DP-1}. Proceeding as in Section \ref{sec:single-rate}, we find that $t^+_\omega$ and $\Delta t_\omega$ are given by the same formulas \eqref{eq:radiation-time} and \eqref{eq:formation-width} as before. Therefore, a mode with longitudinal quasimomentum $p_1$ can emit a photon with frequency $\omega$ only in the vicinity of the time $t_\omega$ at which $2 P^1(t_\omega)v_F = \hbar \omega$, where $P^1(t)=p^1+eEt$ is the instantaneous longitudinal momentum. There is no contribution from $t_\omega^-$. This implies that the spectrum of the radiation emitted by the mode $\vec p$ in $\mathcal{T}$ is restricted to 
\[
0 < \omega < 2(eET-p_1) \frac{v_F}{\hbar} \, .
\]
As a result, the spectrum of the radiation emitted by the electronic gas is limited to the maximum frequency $\omega_{max}=2eETv_F/\hbar$. Moreover, contributions for a given $\omega$ come from quasimomenta such that
\begin{equation}
p_1<eET - \frac{\hbar \omega}{2 v_F} \, .
\label{eq:rad-formation-cutoff}
\end{equation}
More generally, the spectrum in an interval $[t_1,t_2] \subset \mathcal{T}$ is limited to the range of frequencies $\mathcal{F}([t_1,t_2])=[0,2 eEt_2 v_F/\hbar]$, with radiation of frequency $\omega$ being formed by modes in \eqref{eq:finite-duration-cutoff} satisfying the condition: $eEt_1 < p_1 + \hbar \omega /2v_F< eEt_2$. We are assuming, as in Section \ref{sec:single-rate}, the duration of the field to be large compared to the time required for radiation of any frequency to be formed, $T \gg \Delta t_\omega=\sqrt{\hbar/eEv_F}$. As remarked before, the radiation formation width is very narrow for strong electric fields, with $\Delta t_\omega \sim 10^{-14} \text{ s}$ for $|E| \sim 10^6 \text{ V/m}$.

Taking into account the cutoffs \eqref{eq:finite-duration-cutoff} and \eqref{eq:rad-formation-cutoff}, we obtain for the photon emission probability per unit area in the interval $\mathcal{T}$ the formula:
\begin{equation}
\frac{1}{S} \frac{d\Gamma^{(DP)}}{d\Omega \, d\omega} =  \frac{e^2}{\hbar c} \left( \frac{v_F}{c} \right)^2 \frac{\omega}{16 \pi^2 A^4} \Theta([0,\omega_{max}]) \frac{1}{(2\pi \hbar)^2} \int_0^{eET-\frac{\hbar \omega}{2 v_F} } dp_1  \int_{-\infty}^{\infty} dp_2 \, \textrm{e}^{-\pi (\lambda+\lambda')/4} O^2(\vec p, \vec q) \, ,
\label{eq:Gamma-DP-T}
\end{equation}
where $\Theta(I)$ is the characteristic function associated with the interval $I$. Some natural approximations allow for a simplification of this general solution. Firstly, since $T\gg \sqrt{\hbar/eEv_F}$ and the width of the spectrum grows linearly with $T$ according to $\Delta \omega = 2eETv_F/\hbar$, the relevant range of frequencies for the strong field regime lies in the region $\omega \gg \omega_{min}=\sqrt{eEv_F/\hbar}$. From \eqref{eq:A-rho-k}, this corresponds to $|\rho|\gg 1$. Substituting the asymptotic form of the confluent hypergeometric functions \cite{abramowitz},
\begin{multline*}
\frac{\Phi(a,b;z)}{\Gamma(b)} = \frac{e^z z^{a-b}}{\Gamma(a)} \left\{ \sum_{n=0}^{S-1} \frac{(b-a)_n (1-a)_n}{n!} z^{-n} + O\left(|z|^{-S}\right) \right\} \\
+ \frac{\textrm{e}^{-i\pi a} z^{-a}}{\Gamma(b-a)} \left\{ \sum_{n=0}^{R-1} \frac{(a)_n (1+a-b)_n}{n!} (-z)^{-n} + O\left(|z|^{-R}\right)  \right\} \, , \text{ for } -\frac{3\pi}{2} < \arg z < -\frac{\pi}{2} \, ,
\end{multline*}
in the explicit formulas for the $R_{j'j}$ integrals given in Eq.~\eqref{eq:R-integrals}, and keeping only the leading and next-to-leading order terms, we find that, for such large frequencies,
\begin{align}
R_{00}(-\rho) &\simeq  \sqrt{\pi} \textrm{e}^{-i\frac{\pi}{4}} \textrm{e}^{i \rho^2/4} \left| \frac{\rho}{\sqrt{2}} \right|^{-i\frac{\lambda+\lambda'}{2}} \textrm{e}^{-\frac{3\pi}{8}(\lambda+ \lambda')} \left( 1+ i \frac{\lambda \lambda'}{2} \frac{1}{\rho^2} \right) \notag \\
R_{11}(-\rho) &\simeq  \sqrt{\pi} \textrm{e}^{-i\frac{\pi}{4}} \textrm{e}^{i \rho^2/4} \left| \frac{\rho}{\sqrt{2}} \right|^{-i\frac{\lambda+\lambda'}{2}} \textrm{e}^{-\frac{3\pi}{8}(\lambda+ \lambda')} \left( -2i \frac{1}{\rho^2} \right)  \, .
\label{eq:R-asymptotic}
\end{align}
Therefore, we can keep only the term proportional to $\tilde{f}_0$ in \eqref{eq:squared-unpol-DP} as a zeroth order approximation in $\rho^{-1}$. The next order corrections are $\rho^{-2}$ terms proportional to $\tilde{f}_2$. Secondly, we can consider an expansion in the small parameter $v_F/c \simeq 1/300 \ll 1$. Neglecting terms of order $\rho^{-2}$ in the formula for $\tilde{f}_0$, and expanding it to first order in $v_F/c$, we obtain:
\begin{equation}
\tilde{f}_0 \simeq 4  \left(\frac{A \hbar}{v_F}\right)^2 \left[ 1-\sin^2 \theta_\gamma \sin^2 \phi_\gamma - 2 \frac{v_F}{c} \frac{1}{\hbar k} \left( p_2 + \frac{\hbar k}{2} \sin \theta_\gamma \sin \phi_\gamma \right) \right] \, ,
\label{eq:f0-large-rho}
\end{equation}
where we have used the approximations \eqref{eq:low-v-approximation-rho} and \eqref{eq:low-v-approximation-varphi}. We can now substitute \eqref{eq:R-asymptotic} and \eqref{eq:f0-large-rho} in \eqref{eq:Gamma-DP-T} and integrate in $\vec p$ in order to obtain the asymptotic formula:
\begin{multline}
\frac{1}{S} \frac{d\Gamma^{(DP)}}{d\Omega \, d\omega} =  \frac{e^2}{\hbar c} \left( \frac{v_F}{c} \right)^2 \frac{\omega}{2^{9/2} \pi^3 Av_F^2} \left( eET - \frac{\hbar \omega}{2v_F}\right) \Theta([0,\omega_{max}]) \textrm{e}^{-2\pi(mv_F/A)^2} \\ 
(1 - \sin^2 \theta_\gamma \sin^2 \phi_\gamma) \exp\left[ -\frac{\pi}{2} \left(\frac{v_F}{c} \frac{\omega}{\omega_{min}} \sin \theta_\gamma \sin \phi_\gamma \right)^2\right] \, ,
\label{eq:emission-DP}
\end{multline}
valid for $\omega \gg \omega_{min}$. For $eET \gg \hbar \omega/2v_F$, this corresponds to a spectral photon emission rate per unit area:
\begin{multline}
\frac{d\gamma}{d\Omega} =  \frac{e^2}{\hbar c} \left( \frac{v_F}{c} \right)^2 \frac{\omega \sqrt{eE}}{2^{9/2} \pi^3 \sqrt{\hbar} v_F^{3/2}} \textrm{e}^{-2\pi(mv_F/A)^2} \\
\times (1 - \sin^2 \theta_\gamma \sin^2 \phi_\gamma) \exp\left[ -\frac{\pi}{2} \left(\frac{v_F}{c} \frac{\omega}{\omega_{min}} \sin \theta_\gamma \sin \phi_\gamma \right)^2 \right]\, 
\label{eq:emission-rate-DP}
\end{multline}
This formula describes the angular distribution and frequency dependency of photon emission at zero temperature and chemical potential in the strong field regime from a single Dirac particle species.

The photon emission rate \eqref{eq:emission-rate-DP} can be interpreted in a simple manner combining the formula for the free photon emission rate derived in \cite{mecklenburg} with the formula $n=\exp(-\pi \lambda)$ for the probability of electron-hole pair creation by the applied electric field. From our analysis of radiation formation, photons of frequency $\omega$ are emitted at any time $t$ mostly by electrons with longitudinal momentum $P^1 \simeq \hbar \omega/2v_F$. Hence, for $\omega \gg \omega_{min}=\sqrt{eEv_F/\hbar}$, the radiation is produced by electrons with instantaneous momentum $\vec P(t)$ essentially parallel to the applied field, since the transversal component is of order $|p_2| \sim A = \hbar \omega_{min}/v_F$ for quasimomenta which undergo pair creation. Setting $\theta_c=0$ in the free photon emission rate given by Eq.~(13) in \cite{mecklenburg}, putting $n_c=\exp(-\pi \lambda')$ and $n_v=1-\exp(-\pi \lambda)$ to take into account pair creation, and integrating in momentum space, we recover the photon emission rate \eqref{eq:emission-rate-DP} with $m=0$, which is the case studied in that paper. Hence, for frequencies $\omega \gg \omega_{min}$, it is possible to use formulas for free photon emission together with the pair creation probability characteristic of strong electric fields to describe the radiation induced by the external field.  For $\omega \to \omega_{min}$, however, corrections to the leading order approximation of the asymptotic formulas \eqref{eq:R-asymptotic} for the $R$ integrals become relevant, and such procedure is not valid. In this case, one can use the exact expression \eqref{eq:Gamma-DP-T} for numerical studies, or include further $\rho^{-n}$ corrections in \eqref{eq:R-asymptotic} and integrate the corresponding Gaussian integrals in \eqref{eq:Gamma-DP-T}. 

The frequency scale at which the free photon emission rate becomes unreliable is determined by the scale at which the assumption of energy conservation during photon emission breaks down. In fact, in the Fermi's rule approach, a transition with energy scale $\omega$ takes place in a minimum time window $\Delta t \sim 1/\omega$. But the electric field transfers an amount of momentum $\Delta P \sim eE\Delta t$ to the electron during this time, which correspond to an energy $\Delta E \sim v_F \Delta P$ because of the linear dispersion relation. This gives $\Delta E/E \sim (\omega_{min}/\omega)^2$, where $E=\hbar \omega$. Hence, for $\omega \to \omega_{min}$, the energy transferred to the electron during the transition is of the same order as the transition energy itself, and the assumption of energy conservation is not justified. For $\omega \gg \omega_{min}$, energy is approximately conserved, and the free photon emission rate is recovered, as long as one takes into account the effects of pair creation.

\subsection{Conditions for experimental observation}

In the last section, we obtained the unpolarized photon emission rate from graphene in a strong uniform electric field for an initial state with the Fermi level at the Dirac point for a single particle species, which is given by Eq.~\eqref{eq:emission-rate-DP}. The radiation pattern is determined by an interplay between the nonperturbative process of electron-hole creation by the electric background and the interaction of the created pairs with the vacuum of the quantized electromagnetic field, which leads to spontaneous photon emission. The later is treated to first order, which requires electron-photon scattering not to affect considerably the number of created pairs. In this regime, the angular and frequency dependence of the photon emission can be used as a probe of the state of the electron gas, thus providing a means for the observation of the Schwinger effect.

Consequences of nonperturbative electron-hole pair creation for the dc conductivity of graphene were discussed in \cite{rosenstein-82,lewkowicz-84,GGY}. In these works, the duration of the electric field $T$ was interpreted, in the case of ballistic transport in a high-quality sample of length $L$, as the ballistic time $T_{bal}=L/v_F$ required for an electron to cross the graphene sample. In the presence of impurities, $T$ is interpreted as a typical time of flight $T_{fl}$ between collisions. In both cases, the dc conductivity becomes proportional to $E^{3/2}$, which characterizes the so-called superlinear regime. Here we adopt the same interpretation for $T$ in order to translate results obtained for an infinite plane to the realistic case of a finite sample of length $L$. For more details on this, see the discussion in Section IV of \cite{GGY}. Moreover, we assume that an experimental realization of the charge neutrality point is available. According to \cite{mayorov}, the Dirac point can be approached experimentally within $1 \text{ meV}$.

According to the discussion in the last section, the spectral photon emission rate \eqref{eq:emission-rate-DP} is valid in the range of frequencies $\omega_{min} < \omega < \omega_{max}$, with
\begin{equation}
\omega_{min} = \sqrt{\frac{eEv_F}{\hbar}} \, , \qquad
\omega_{max} = \frac{2eE T v_F}{\hbar} \, .
\label{eq:frequency-window}
\end{equation}
The lower bound is required for the validity of the asymptotic approximations \eqref{eq:R-asymptotic} and \eqref{eq:f0-large-rho}; as $\omega \to \omega_{min}$, the emission rate \eqref{eq:emission-rate-DP} acquires higher-order $\rho^{-n}$ corrections, where $\rho=\omega/\omega_{min}$. The upper bound $\omega_{max}$  accounts for the fact that the production of high-frequency radiation requires the acceleration of the created pairs to sufficiently large momenta, which is limited by the size of the sample. These frequency bounds for the observation of the Schwinger effect agree with those obtained in \cite{lewkowicz-84}. For concreteness, set $L \sim 1 \,\mu\text{m}$. This implies a ballistic time $T \sim 10^{-12} \text{ s}$.  For an applied field of $\sim 10^6 \text{ V/m}$ and ballistic transport, we find $\nu_{min} \simeq 6 \text{ THz}$ and $\nu_{max} \simeq 484 \text{ THz}$, and the radiation spectrum ranges from the infrared to the lower end of the visible.  In the presence of impurities, the upper bound is lowered according to \eqref{eq:frequency-window}, with $T$ given by the typical time of flight. For an applied field of $\sim 10^4 \text{ V/m}$, in comparison, the spectrum ranges from $0.6$ to $4.8\text{ THz}$, and lies entirely in the far-infrared.

The photon emission rate \eqref{eq:emission-rate-DP} refers to a single Dirac point. To the approximation considered, the emission rate does not depend on which Dirac point is considered, and in order to account for (true) spin and valley degeneracy, we multiply \eqref{eq:emission-rate-DP} by four. In the massless case, we obtain:
\begin{equation}
\frac{d\gamma}{d\Omega} =  \tau \omega \frac{\sqrt{eE}}{\sqrt{\hbar}v_F^{3/2}} (1 - \sin^2 \theta_\gamma \sin^2 \phi_\gamma) \exp\left[ -\frac{\pi}{2} \left(\frac{v_F}{c} \frac{\omega}{\omega_{min}} \sin \theta_\gamma \sin \phi_\gamma \right)^2 \right] \, ,
\label{eq:emission-rate-DP-graphene}
\end{equation}
where $\tau=\alpha (v_F/c)^2 2^{-5/2} \pi^{-3} \simeq 4.6 \times 10^{-10}$, with $\alpha = e^2/\hbar c$ the fine structure constant, is completely determined by fundamental constants. For $\omega/\omega_{min} \ll c/v_F$, the exponential attenuation can be neglected, and the angular distribution has the simple form $1 - \sin^2 \theta_\gamma \sin^2 \phi_\gamma$, increases linearly with $\omega$, and scales as $\sqrt{E}$. This fractional power dependence on the applied field is a hallmark of the nonperturbative creation of electron-hole pairs, and its observation would provide evidence for the Schwinger mechanism. At these frequencies, the angular distribution is not sensitive to the details of the distribution of pairs in the transversal momentum $p_2$. In fact, the same emission rate is found for any distribution of pairs such that $\int dp_2 n^2(p_2)=A/\sqrt{2}$, where $n(p_2)$ is the number of electron-hole pairs with transversal momentum $p_2$, assuming negligible momentum transfer, $\vec p \simeq \vec q$, which is a good approximation in this regime. For the Schwinger distribution, $n(p_2)=\text{e}^{-\pi p_2^2/A^2}$.

When $\omega/\omega_{min}$ is comparable to $c/v_F$, the photon emission is progressively focused along the vertical plane parallel to the applied field, since the transversal in-plane component $k_2$ of the photon wavenumber is restricted to the region $\hbar k_2 \sim A$, for any $k$. This is a direct consequence of the Gaussian pattern $n(p_2)=\text{e}^{-\pi p_2^2/A^2}$ of the distribution of created pairs along the transversal electron quasimomentum axis. In fact, since holes are created only for $p_2 \sim A$, photon emission transitions with large transversal momentum transfer are blocked by the Pauli exclusion principle, leading to the specific attenuation pattern displayed in \eqref{eq:emission-rate-DP-graphene}. The observation of such focusing of the angular distribution of photon emission for high-frequency radiation would be a clear indication of the Schwinger mechanism in graphene. The magnitude of this effect is determined by the coefficient:
\[
f=\frac{\omega_{max}}{\omega_{min}} = 2 \sqrt{\frac{eUL}{\hbar v_F}} \, ,
\]
where $U$ is the voltage applied to the sample. The exponential attenuation in \eqref{eq:emission-rate-DP-graphene} becomes relevant around $\omega/\omega_{min} \sim 100$. As an illustration, for $f = 50$, the factor $\exp[-\pi(\omega v_F/\omega_{min}c )^2/2]$ is equal to $0.96$, falling to $0.50$ at $f = 200$, and reaching $0.21$ at $f=300$. This regime is achieved for $UL \sim 1.6 \text{ V$\mu$m}$ (for $f=100$). Therefore, the observation of the photon emission rate \eqref{eq:emission-rate-DP-graphene} in full detail would require the achievement of ballistic transport in samples of a few microns at the Dirac point with an applied voltage of a few volts, and would occur in the near-infrared spectrum. For $L=5 \text{ $\mu$m}$ and $U=1 \text{ V}$, for instance, considering the focusing of the radiation significant (attenuation larger than $\sim 4\%$) for frequencies higher than $\sim 50 \, \omega_{min}$, we have the threshold $\nu_0  \sim 140 \text{ THz}$ for the observation of the exponential factor in \eqref{eq:emission-rate-DP-graphene}.


\section{Conclusion}
\label{sec:conclusion}

We have developed a comprehensive theoretical approach to the calculation of photon emission probabilities from many-particle states in graphene in the presence of a strong uniform electric field. The treatment is nonperturbative in the electric background, and applies in situations where it is strong enough to induce nonlinear tunneling between energy bands. Our main result was the calculation of the spectral photon emission rate induced by the applied field for an initial state with Fermi level at the Dirac point. Such state corresponds to the vacuum of QED in the Dirac model, and the pattern of photon emission exhibits distinctive features which can be traced to the Schwinger effect. In particular, the photon emission rate scales with the electric field $E$ as $\sqrt{|E|}$, reflecting the nonperturbative character of the pair creation process, and displays a characteristic focusing in the plane perpendicular to the graphene sheet and parallel to the applied field for frequencies $\omega \gg \omega_{min}=\sqrt{eEv_F/\hbar}$. This focusing occurs in the near-infrared for clean micrometer-scale samples with an applied voltage of a few volts, and constitutes a clear indication of the Schwinger mechanism in graphene.

Photon emission due to the interaction with the quantized $3D$ electromagnetic field has been previously studied in free space \cite{mecklenburg}, and our results extend that work to the case of an applied dc voltage. Processes which are forbidden by energy and momentum conservation in the free case are allowed in the presence of the external field, and lead to quantum interference terms in the probabilities of photon emission from many-electron states, which we take fully into account. Moreover, the photon emission rate cannot be calculated using the standard Fermi's Golden Rule approach, since energy is not conserved. Instead, we computed total transition amplitudes (integrated in space and time) of radiative first-order processes using exact solutions of the Dirac equation in the electric background, from which photon emission rates were extracted through an analysis of the process of radiation formation in the strong field regime characterized by ballistic times $t_{bal} \gg \Delta t_{\omega}=\sqrt{\hbar/eEv_F}$. This method was adapted from classical works on radiative processes in a constant electric field \cite{nikishov-69,nikishov-70,nikishov-71} developed in the context of QED with unstable vacuum \cite{FGS}. Our work provides a basis for the rigorous analysis of the dc conductivity of pristine graphene at low temperature as limited by the interaction with the vacuum of the photon field.

In the case of a single particle in graphene, the photon emission is strongly time-dependent, with radiation with frequency $\omega$ being emitted only in a time window of width $\Delta t_{\omega}$ around the times $t^{\pm}_\omega$ at which $2 P^1\left(t^{\pm}_\omega \right)v_F = \pm \hbar \omega$, where $P^1(t)=p^1+eEt$ is the component of the electron momentum parallel to the field. The angular distribution has corrections of order $v_F/c$ as compared to the free case. At the Dirac point, the photon emission rate for any $\omega > \omega_{min}$ approaches a constant value for $T \gg \hbar \omega/ 2eEv_F$, where $T$ is the duration of the applied field. For large frequencies $\omega \gg \omega_{min}$, the spectral photon emission rate can be well approximated by a Fermi rule approach using free photon emission amplitudes and a time-dependent number of pairs due to pair creation, as done in \cite{lewkowicz-84}, for instance. As $\omega$ approaches $\omega_{min}$, large $(\omega/\omega_{min})^{-n}$ corrections become relevant, and it is crucial to use the exact solutions of the Dirac equation in the calculation of the amplitudes of the relevant Feynman diagrams.

\begin{acknowledgments}
NY acknowledges support from CAPES (PRODOC program).
\end{acknowledgments}

\appendix

\section{Calculation of the time integrals}

\subsection{J and I integrals}

The $J_{j'j}$ integrals are defined in \eqref{eq:nikishov-J}. It is enough to consider the simpler integrals $I_{j'j}$, due to Eq.~\eqref{eq:J-I}. An explicit expression for the $I_{00}$ integral can be found in Eq.~(A.12) at the Appendix of \cite{nikishov-71}. In order to translate that result to our notation, it is necessary to apply the substitutions $\nu \to \beta$ and $\zeta \to -i \rho^2/2$. Formulae for the $I_{j'j}$ integrals with $j,j' \neq 0$ are obtained from that for $I_{00}$ with the substitution
\[
\lambda \to \lambda - 2ji \,  , \qquad \lambda' \to \lambda' - 2 j' i \, ,
\]
leading to the expression displayed in Eq.~\eqref{eq:I-integrals}. The formulas for the $I_{j'j}$ involve confluent hypergeometric functions whose arguments differ by small integers. Simple algebraic identities called contiguity relations hold among such hypergeometric functions (see \cite{erdelyi}), and can be translated into identities among the $I_{j'j}$ integrals. In fact, the contiguity relations:
\begin{gather*}
(1+i \nu) \Phi\left( 1+i \frac{\lambda}{2}, 1+ i \nu \right) - (1-i \nu) \Phi\left(i \frac{\lambda}{2}, 1+ i\nu \right) + i \frac{\rho^2}{2} \Phi\left( 1 + i \frac{\lambda}{2}, 2+ i \nu \right) = 0  \, , \\
i \frac{\lambda}{2} \Phi\left( i \frac{\lambda'}{2}, 1- i \nu \right) - i \frac{\lambda'}{2} \Phi\left( 1+i \frac{\lambda'}{2}, 1- i \nu \right) -  i \nu \Phi\left( i \frac{\lambda'}{2}, -i \nu \right) = 0 \, ,
\end{gather*}
(the dependence on $-i \rho^2/2$ was omitted) together with the explicit formulas for $I_{00}$, $I_{01}$ and $I_{11}$ given in \eqref{eq:I-integrals} lead to the first identity in \eqref{eq:contiguity-relations}. The second identity is obtained from the relations:
\begin{gather*}
(1-i\nu) \Phi\left( 1+ i \frac{\lambda'}{2}, 1-i\nu \right) - (1-i\nu)  \Phi\left( i\frac{\lambda'}{2}, 1-i\nu \right) + i \frac{\rho^2}{2} \Phi\left(1+i\frac{\lambda'}{2},2-i\nu \right) = 0 \, , \\
i\frac{\lambda'}{2} \Phi \left( i\frac{\lambda}{2}, 1+ i \nu \right) - i \frac{\lambda}{2} \Phi\left(1+i \frac{\lambda}{2},1+i\nu \right) + i \nu \Phi\left( i \frac{\lambda}{2},i \nu \right) = 0 \, ,
\end{gather*}
and the explicit formulas for $I_{00}$, $I_{10}$ and $I_{11}$.

\subsection{L and K integrals}

The techniques used in \cite{nikishov-71} for the calculation of the $J_{j'j'}$ can be applied to the calculation of all time integrals found in the main text. In this section, we consider the example of the $L_{j'j}$ integrals in full detail. We show all steps involved in the calculation of these integrals, so that this section can be read as a short didactic exposition of the general procedure introduced in \cite{nikishov-71}. The same steps are involved in the calculation of all time integrals describing first order radiative processes in the presence of a constant electric field.

The $L_{j'j}$ integrals are defined as
\begin{equation}
L_{j'j} = \int_{-\infty}^{+ \infty} dt \, D_{-i \lambda'/2-j'}[-(1+i)\xi'] D_{i \lambda/2-j}[(1-i)\xi] \textrm{e}^{-i \omega t}\, .
\label{eq:L-integral-app}
\end{equation}
Introducing a new variable $v$ and a parameter $v_1$ through the transformation:
\begin{equation}
-\xi' = v + \frac{v_1}{2} \, , \qquad 
-\xi = v - \frac{v_1}{2} \, ,
\label{eq:xi-v}
\end{equation}
which leads to
\begin{equation}
t = - \sqrt{\frac{\hbar}{eEv_F}} v + \frac{p_1 + q_1}{2eE} \, , \qquad dt = - \sqrt{\frac{\hbar}{eEv_F}} dv \, ,
\label{eq:t-to-v}
\end{equation}
the original integral transforms into
\begin{align*}
& L_{j'j} = \frac{\hbar}{A v_F} \exp \left(- i \omega \frac{p_1 + q_1}{2eE}  \right) K_{j'j}(v_0,v_1) \, , \\
& K_{j'j}(v_0,v_1) = \int_{-\infty}^{+ \infty} dv \, D_{-i \lambda'/2-j'}\left[(1+i) \left(v+\frac{v_1}{2}  \right)\right] D_{i \lambda/2-j}\left[-(1-i)\left(v-\frac{v_1}{2}  \right)\right] \textrm{e}^{-i v_0 v}\, ,
\end{align*}
where $v_0=-\hbar \omega/ A v_F$.

The integrals $K_{j'j}(v_0,v_1)$ have the general form:
\begin{equation}
\mathcal{J}(v_0,v_1) = \int_{-\infty}^{+ \infty} dv \, f_{\Lambda'}(v +v_1/2) f_\Lambda(v-v_1/2) \textrm{e}^{- i v_0 v} \, ,
\label{eq:boost-integral}
\end{equation}
where the functions $f_\Lambda(x)$ satisfy the differential equation
\begin{equation}
\left( \frac{d^2}{dv^2} + v^2 + \Lambda \right) f_\Lambda(v) = 0 \, ,
\label{eq:fnabla-def}
\end{equation}
with
\begin{equation}
\Lambda = \lambda + 2 ji - i \, , \qquad 
\Lambda' = \lambda' - 2 j'i+ i \, .
\label{k-lambda-lambda}
\end{equation}
Introducing hyperbolic coordinates
\[
v_0 = \rho \cosh \varphi \, , \qquad v_1 = \rho \sinh \varphi \, ,
\]
and taking the derivative of the integral $\mathcal{J}(\rho,\varphi)$ with respect to $\varphi$, we obtain the first-order differential equation
\[
\frac{\partial \mathcal{J}}{\partial \varphi}(\rho,\varphi) = i \frac{\Lambda^\prime - \Lambda}{2} \mathcal{J}(\rho,\varphi) \, ,
\]
which is solved by
\begin{equation}
\mathcal{J}(\rho,\varphi) = \exp\left(i \frac{\Lambda^\prime - \Lambda}{2} \varphi \right) \mathcal{J}(\rho,0) \, .
\label{eq:boost}
\end{equation}
This reduces the problem of integrating \eqref{eq:boost-integral} to the calculation of the simpler integral:
\begin{equation}
I(\rho) = \mathcal{J}(\rho,0) = \int_{-\infty}^{+ \infty} dv \, f_{\Lambda'}(v) f_{\Lambda}(v) \textrm{e}^{- i \rho v} \, ,
\label{eq:I-rho-integral}
\end{equation}
where $\rho = - \sqrt{v_0^2 - v_3^2}$.

In order to compute the integral $I(\rho)$, we first show that it satisfies a certain differential equation, and then solve this equation with appropriate boundary conditions. It can be checked that the integral $I(\rho)$ satisfies:
\begin{equation}
\left[ \frac{d^2}{d \rho^2} + \frac{1}{\rho} \frac{d}{d \rho} + \frac{(\Lambda' - \Lambda)^2}{4 \rho^2} I + \frac{\rho^2}{4} - \frac{\Lambda + \Lambda'}{2}  \right] I(\rho) = 0 \, .
\label{eq:I-rho-eqdiff}
\end{equation}
Introducing the variables
\[
\zeta = - i \frac{\rho^2}{2} \, , \qquad \mu = i \frac{\Lambda - \Lambda'}{4} \, , \qquad \alpha = - i \frac{\Lambda + \Lambda'}{4} \, ,
\]
it follows that the function
\begin{equation}
F(\zeta) = \textrm{e}^{-\frac{\zeta}{2}} \zeta^{\frac{1}{2}} I(\zeta) \, ,
\label{eq:F-df}
\end{equation}
satisfies a confluent hypergeometric equation,
\[
\frac{d^2 F}{d\zeta^2} + \frac{dF}{d\zeta} + \left( \frac{\alpha}{2} + \frac{\frac{1}{4}- \mu^2}{\zeta^2}  \right) F = 0 \, .
\]
Linearly independent solutions of the confluent hypergeometric equation are known, and we conclude that $I(\rho)$ must have the form:
\begin{equation}
I(\rho) = \textrm{e}^{- \frac{\zeta}{2}} \left[ C_1 \zeta^{i \frac{\Lambda- \Lambda'}{4}} \Phi \left(\frac{1}{2} + i \frac{\Lambda}{2}, 1+ i\frac{\Lambda- \Lambda'}{2};\zeta \right) + C_2 \zeta^{- i \frac{\Lambda- \Lambda'}{4}} \Phi\left(\frac{1}{2} + i \frac{\Lambda'}{2},1- i \frac{\Lambda- \Lambda'}{2}; \zeta  \right) \right] \, ,
\label{eq:solution-I-rho}
\end{equation}
where the $C_i$ are undetermined coefficients, which must be fixed by appropriate boundary conditions so that the solution corresponds to the original integral.

According to the discussion up to this point, the original integral \eqref{eq:L-integral-app} can be written as:
\begin{align}
& L_{j'j} = \frac{\hbar}{A v_F} \exp\left( - i \omega \frac{p_1 + q_1}{2eE} \right) \textrm{e}^{-i \beta \varphi} \textrm{e}^{(j + j'-1)\varphi} K_{j'j}(\rho) \, ,  \nonumber \\
& K_{j'j}(\rho) = \int_{-\infty}^{+\infty} dv D_{-i\lambda'/2 - j'}[(1+i)v] D_{i\lambda/2-j}[-(1-i)v] \textrm{e}^{- i \rho v} \, ,
\label{eq:L-K-app}
\end{align}
where $K_{j'j}$ has the form \eqref{eq:solution-I-rho}. In particular, from \eqref{k-lambda-lambda} and \eqref{eq:solution-I-rho},
\begin{multline}
K_{00}(\rho) = \textrm{e}^{i \rho^2/4} \left[ C_1 \, \left(-i\frac{\rho^2}{2}\right)^{\frac{1}{2} +i \frac{\beta}{2}} \Phi \left(1+ i \frac{\lambda}{2}, 2+ i \beta;-i\frac{\rho^2}{2} \right) \right. \\
	\left. + C_2 \, \left(-i\frac{\rho^2}{2}\right)^{-\frac{1}{2} - i \frac{\beta}{2}} \Phi\left(i \frac{\lambda'}{2}, - i \beta; -i\frac{\rho^2}{2}  \right) \right] \, .
\label{eq:k-solution-superposition}
\end{multline}
The coefficients $C_i$ are fixed by an analysis of the $\rho \to 0$ limit. In this limit, only large $|v|$ contribute to the integral, and we can use the asymptotic form of the Weber functions $D_\nu$ in order to integrate \eqref{eq:L-K-app}.

We decompose $K_{00}$ into a sum of contributions from the positive and negative semi-axes,
\begin{gather}
K_{00}(\rho) = K^+_{00}(\rho) - K^-_{00}(\rho) \, , \nonumber \\
K^+_{00}(\rho) = \int_0^{+ \infty} dv \, D_{-i\lambda'/2}[(1+i)v] D_{i\lambda/2}[-(1-i)v] \textrm{e}^{- i \rho v} \, , 
\label{eq:k00+} \\
K^-_{00}(\rho) = \int_{- \infty}^{0} dv \, D_{-i\lambda'/2}[(1+i)v] D_{i\lambda/2}[-(1-i)v] \textrm{e}^{- i \rho v} \, .
\label{eq:k00-}
\end{gather}
Consider the contribution $K^+_{00}(\rho)$. From the general identity (see \cite{erdelyi})
\[
D_\nu(z) = \textrm{e}^{\nu \pi i} D_\nu(-z) + \frac{\sqrt{2 \pi}}{\Gamma(-\nu)} \textrm{e}^{i (\nu+1)\pi/2 } D_{-\nu-1}(-iz) \, , 
\]
we find that
\[
D_{i\lambda/2}[-(1-i)v] = \textrm{e}^{-\pi \lambda/2} D_{i\lambda/2}[(1-i)v] + \frac{\sqrt{2\pi}}{\Gamma(-i \lambda/2)} \textrm{e}^{-\pi \lambda/4} \textrm{e}^{i\pi/2} D_{-i\lambda/2-1}[(1+i)v] \, .
\]
Substituting this expression in \eqref{eq:k00+}, and using the asymptotic approximation
\begin{equation}
D_\nu(z) \simeq z^\nu \textrm{e}^{-z^2} \, ,
\label{eq:weber-asymptotic}
\end{equation}
which is valid for $|\nu/z|\ll 1$ and $|\arg z|<3\pi/4$, we obtain
\begin{align*}
K_{00}^+ & \simeq 2^{i \beta/2} \textrm{e}^{\pi \lambda'/8} \textrm{e}^{-3 \pi \lambda/8} \int_0^\infty dv \, v^{i \beta} \textrm{e}^{- i \rho v} \, \\
		& \simeq - \frac{1}{\sqrt{2}} \textrm{e}^{5\pi \lambda'/8} \textrm{e}^{-7 \pi \lambda/8} \left(\frac{-i\rho}{\sqrt{2}} \right)^{-1-i\beta} \Gamma(1+i\beta) \, ,
\end{align*}
where a term proportional to $\exp(-iv^2)$ was neglected in the integrand. The integral of such term is dominated by a region close to $v=0$, and does not contribute to the limit $\rho \to 0$. Now consider the contribution $K^-_{00}(\rho)$. From the identity \cite{erdelyi}
\[
D_\nu(z) = \textrm{e}^{-\nu \pi i} D_\nu(-z) + \frac{\sqrt{2 \pi}}{\Gamma(-\nu)} \textrm{e}^{-i (\nu+1)\pi/2 } D_{-\nu-1}(iz) \, , 
\]
we obtain
\[
D_{-i\lambda'/2}[(1+i)v] = \textrm{e}^{-\pi \lambda'/2} D_{-i\lambda'/2}[-(1+i)v] + \frac{\sqrt{2\pi}}{\Gamma(i \lambda'/2)} \textrm{e}^{-\pi \lambda'/4} \textrm{e}^{-i \pi/2} D_{i\lambda'/2-1}[-(1-i)v] \, .
\]
Substituting this expression in \eqref{eq:k00-}, using the asymptotic formula \eqref{eq:weber-asymptotic}, and neglecting a term proportional to $\exp(iv^2)$ in the integrand, we obtain:
\begin{align*}
K_{00}^- & \simeq 2^{i \beta/2}  \textrm{e}^{\pi \lambda/8} \textrm{e}^{-3 \pi \lambda'/8} \int_{-\infty}^0 dv \, (-v)^{i\beta} \textrm{e}^{- i \rho v} \\
	& \simeq \frac{1}{\sqrt{2}}  \textrm{e}^{\pi \lambda/8} \textrm{e}^{-3 \pi \lambda'/8} \left(\frac{-i\rho}{\sqrt{2}} \right)^{-1-i\beta} \Gamma(1+i\beta) \, .
\end{align*}
Summing both contributions,
\[
K_{00}(\rho) \simeq \frac{1}{\sqrt{2}} \Gamma(1+i\beta) \textrm{e}^{\pi \lambda/8} \textrm{e}^{-3 \pi \lambda'/8} \left(\frac{-i\rho}{\sqrt{2}} \right)^{-1-i\beta} \left(1-\textrm{e}^{\pi \lambda - \pi \lambda'}  \right) \, .
\]

Comparing this result with the $\rho \to 0$ behavior of the expression \eqref{eq:k-solution-superposition}, we find that
\[
C_1 =0 \, , \qquad C_2 = \frac{1}{\sqrt{2}} \Gamma(1+i\beta)  \textrm{e}^{\pi \lambda/2} \textrm{e}^{-3 \pi \lambda'/4} \left(1-\textrm{e}^{\pi \lambda - \pi \lambda'}  \right) \, ,
\]
so that the final solution is
\[
K_{00}(\rho) = \frac{\textrm{e}^{i \rho^2/4}}{\sqrt{2}} \Gamma(1+i\beta)  \, \textrm{e}^{\pi \lambda/8 -3 \pi \lambda'/8} \left(1-\textrm{e}^{\pi \lambda - \pi \lambda'} \right)  \left(\frac{-i\rho}{\sqrt{2}} \right)^{-1-i\beta} \Phi\left(i \frac{\lambda'}{2}, - i \beta; -i\frac{\rho^2}{2}  \right) \, .
\]
The $K_{j'j}$ integrals with indices different from zero are obtained with the substitution
\[
\lambda \to \lambda + 2 j i \, , \qquad \lambda' \to \lambda' - 2 j' i \, ,
\]
which leads to:
\begin{multline}
K_{j'j}(\rho) = \frac{\textrm{e}^{i \rho^2/4}}{\sqrt{2}} \Gamma(1+i\beta-j-j') \textrm{e}^{\pi \lambda/8 -3 \pi \lambda'/8} \textrm{e}^{i(j + 3j' )\pi/4} \\
\times \left(1-\textrm{e}^{\pi \lambda - \pi \lambda'}\right) \left(\frac{-i\rho}{\sqrt{2}} \right)^{-1-i\beta+j+j'}  \Phi\left(i \frac{\lambda'}{2}+j',-i \beta+j+j'; -i\frac{\rho^2}{2}  \right) \, .
\label{eq:k-solution}
\end{multline}
Plugging this expression in Eq.~\eqref{eq:L-K-app} gives the exact solution of the integral $L_{j'j}$.

The expressions for the distinct $K_{j'j}$ involve confluent hypergeometric functions whose arguments differ by small integers. Contiguity relations holding among such hypergeometric functions translate into identities among the $K_{j'j}$ integrals. In fact, the identity
\[
i\frac{\lambda}{2} \Phi\left(i \frac{\lambda'}{2},1-i \beta \right) - i\frac{\lambda'}{2} \Phi\left(1+i \frac{\lambda'}{2},1-i \beta \right) - i\beta \Phi\left(i \frac{\lambda'}{2},-i \beta \right) = 0 \, ,
\]
corresponds to the first identity in \eqref{eq:contiguity-relations-K},
while the contiguity relation
\[
(1-i\beta) \Phi\left(1+i \frac{\lambda'}{2},1-i\beta\right) - (1-i\beta) \Phi\left(i \frac{\lambda'}{2},1-i \beta \right) + \frac{i \rho^2}{2} \Phi\left(1+i \frac{\lambda'}{2},2-i \beta \right) = 0 \, ,
\]
corresponds to the second identity in \eqref{eq:contiguity-relations-K}.

\subsection{S and R integrals}

The $S_{j'j}$ integrals are defined by:
\begin{equation}
S_{j'j}(\omega) = \int_{-\infty}^{+ \infty} dt \, D_{-i \lambda'/2-j'}[-(1+i)\xi'] D_{-i \lambda/2-j}[-(1+i)\xi] \textrm{e}^{-i \omega t}\, .
\end{equation}
Applying the transformation \eqref{eq:xi-v}, these reduce to
\begin{align*}
&S_{j'j} = \frac{\hbar}{A v_F} \exp \left(- i \omega \frac{p_1 + q_1}{2eE}  \right) R_{j'j}(v_0,v_1) \\
&R_{j'j}(v_0,v_1) = \int_{-\infty}^{+ \infty} dv \, D_{-i \lambda'/2-j'}\left[(1+i) \left(v+\frac{v_1}{2}  \right)\right] D_{-i \lambda/2-j}\left[(1+i)\left(v-\frac{v_1}{2}  \right)\right] \textrm{e}^{-i v_0 v}\, .
\end{align*}
The integral $R_{j'j}(v_0,v_1)$ has the general form \eqref{eq:boost-integral}, with
\begin{align*}
& \Lambda = \lambda - 2 ji+ i \, , \\
& \Lambda' = \lambda' - 2 j'i+ i \, .
\end{align*}
This allows us to factor out the dependence on the longitudinal momenta $p_1,q_1$ using \eqref{eq:boost},
\[
R_{j'j}(v_0,v_1) =  \textrm{e}^{(j'-j)\varphi} \textrm{e}^{-i\beta \varphi} R_{j'j}(\rho) \, ,
\]
where $R_{j'j}(\rho)$ has the form \eqref{eq:solution-I-rho}. The coefficients $C_i$ are fixed by the limit $\rho \to 0$ for each $R_{j'j}(\rho)$.  It is enough to compute $R_{00}$, since the remaining integrals are obtained applying the substitutions
\[
\lambda \to \lambda - 2ji \, , \qquad \lambda' \to \lambda' - 2j'i \, ,
\]
in the formula for $R_{00}(\rho)$. We obtain in this way the formula in Eq.~\eqref{eq:R-integrals}.

The following contiguity relations of confluent hypergeometric functions:
\begin{gather*}
(1+i\beta) \Phi\left(1+i \frac{\lambda}{2},1+i\beta \right) - (1+i\beta) \Phi\left(i \frac{\lambda}{2},1+i\beta \right) + \frac{i \rho^2}{2} \Phi\left(1+i \frac{\lambda}{2},2+i\beta \right) = 0 \, , 
\\
i\frac{\lambda}{2} \Phi\left(i \frac{\lambda'}{2},1-i\beta \right) - i\frac{\lambda'}{2}  \Phi\left(1+i \frac{\lambda'}{2},1-i\beta \right) -i\beta  \Phi\left(i \frac{\lambda'}{2},i\beta \right) =0 \, ,
\end{gather*}
together with the explicit formulas for $R_{00}$, $R_{01}$ and $R_{11}$ given in Eq.~\eqref{eq:R-integrals} lead to the first relation among $R_{j'j'}$ integrals in \eqref{eq:R-contiguity}. The contiguity relations
\begin{gather*}
i\frac{\lambda'}{2} \Phi\left(i \frac{\lambda}{2},1+i\beta \right) - i\frac{\lambda}{2} \Phi\left(1+i \frac{\lambda}{2},1+i\beta \right) + i\beta \Phi\left(i\frac{\lambda}{2},i\beta \right) = 0 \, , 
\\
(1-i\beta) \Phi\left(1+i \frac{\lambda'}{2},1-i\beta \right) - (1-i\beta) \Phi\left(i \frac{\lambda'}{2},1-i\beta \right) + \frac{i \rho^2}{2} \Phi\left(1+i \frac{\lambda'}{2},2-i\beta \right) = 0 \, ,
\end{gather*}
together with the explicit formulas for $R_{00}$, $R_{10}$ and $R_{11}$ given in Eq.~\eqref{eq:R-integrals} lead to the second relation in \eqref{eq:R-contiguity}.

\end{document}